\documentclass{article}

\usepackage{graphicx}
\usepackage{multirow} 
\usepackage{balance}
\usepackage{a4,amsmath}
\usepackage{amssymb}
\usepackage{listings} 
\include{epsf}

\def\S{\mbox{\large $\rhd\!\!\!\lhd$}}

\newcommand{\pf}{\paragraph{Proof}}

\newtheorem{definition}{Definition}

\newtheorem{proposition}{Proposition}

\newtheorem{algorithm}{Algorithm}

\title{Storage Workload Modelling by Hidden Markov Models: Application to FLASH Memory}

\author{
P. G. Harrison$^1$ S. K. Harrison$^2$ N. M. Patel$^3$ S. Zertal$^4$ \\ ~ \\
\small
$^1$  Imperial College London, South Kensington Campus, London SW7 2AZ,  
UK \\
\small {\tt  pgh@doc.ic.ac.uk} \\ ~ \\
\small 
$^2$  Royal Holloway University London, Egham,  Middlesex,
UK \\
\small {\tt  S.K.Harrison@rhul.ac.uk} \\ ~ \\
\small $^3$ NetApp, Inc., 495 East Java Drive, Sunnyvale, CA 94089, USA \\ \small  {\tt  
naresh@netapp.com}\\ ~ \\
\small $^4$ PRiSM, Universit\'e  de Versailles, 45, Av. des
Etats-Unis, 78000  Versailles, France \\ \small  {\tt  
Zertal@prism.uvsq.fr}
}

\begin{document}

\maketitle

\begin{abstract}
A workload analysis technique is presented that processes data from
operation type traces and creates a Hidden Markov Model (HMM) to
represent the workload that generated those traces.  The HMM can be
used to create representative traces for performance models, such as
simulators, avoiding the need to repeatedly acquire suitable traces.
It can also be used to estimate directly the transition probabilities and rates of a Markov modulated arrival process, for use as input to an analytical performance model of Flash memory.  The HMMs obtained from industrial workloads -- both synthetic benchmarks, preprocessed by a file translation layer, and real, time-stamped user traces -- are validated by comparing their autocorrelation functions and other statistics with those of the corresponding monitored time series.  Further, the performance model applications, referred to above, are illustrated by numerical examples.  
\end{abstract}

\section{Introduction}

Networked storage has always had to deal with traffic from many servers while at the same time providing adequate storage performance to all applications. The recent trend of server virtualization consolidates many applications into a smaller number of server systems and much of the traffic to shared storage remains unchanged. However, data management operations and tighter integration between virtualized systems and shared storage increase the variety of traffic flows in the mix during certain time periods. Oversubscribing the shared resources such as disks is a problem, especially in cloud service environments where strict performance service level agreements need to be met and true performance isolation with independent hardware is too costly. In this type of environment the workload needs to represent time-varying correlated traffic streams that may create different resource bottlenecks in the system at different times. This becomes particularly important as contemporary storage systems introduce a large layer of Flash in the memory hierarchy between DRAM and hard disk drives.   This leads to very different performance characteristics and adds to the complexity of storage systems as potential bottlenecks may develop in any of the layers at different times.  This further underlines the need to represent time-varying, correlated traffic streams in workload models.

Flash storage devices are ubiquitous in the consumer space and
quietly establishing a foothold in the enterprise space as Flash-based
PCIe cards or as solid-state drives. Flash devices can be used in file
service applications (home directories, etc.) to reduce the disk spindle
requirements substantially. For database applications, Flash is used in
on-line transaction processing (OLTP), decision support systems (DSS),
and for backup/recovery, etc. Often, a burst of activity is generated
from maintenance tasks such as loading databases and creating indexes.
In general, OLTP applications benefit the most from Flash because small
random reads to index tables are common and Flash provides low latency
read access. DSS applications tend to have larger and mostly sequential
reads and this reduces the advantage of Flash over disks.

\subsection{Storage workload models}
Storage workload characterisation is important for all kinds of performance evaluation: measurement, simulation and analytical modelling. For measurement and simulation, IO traces from production systems can be used but they are often hard to obtain, require a lot of space, and tend to be very old. The large sizes of the traces (many Gigabytes) create a barrier to broad usage as download times are long and resource-intensive. An alternative to this would be to construct a compact representation (i.e. with a small number of key parameters) of a class of traces that could be used to create typical traces that preserve the key characteristics.  Indeed the workload model could then be used to experiment with new storage system designs, especially for Flash or hybrid Flash and disk. Moreover, owners of production systems may be more open to providing key characteristics than long traces of their applications. 
Some way to extract representative workload parameters from traces, or from production systems, would be highly desirable and this is the key issue that we explore in this paper in the context of storage.
	

\subsection{Terminology: Disks vs SSDs/Flash}
Traditional storage in enterprise systems has come in two flavours of hard disk drives (HDD): high-speed storage (Fibre Channel or SAS HDD) and high-capacity storage (SATA HDD). Over the last few years, Flash-based storage or solid-state drive (SSD) technology has emerged as an ultra-high speed storage class that threatens to make obsolete the high-speed HDDs provided relative cost reduction curves are sustainable. For enterprise storage, the SSDs contain single-level cell (SLC) NAND Flash, although some SSDs use the lower cost multi-level cell (MLC) NAND Flash, despite its inherently faster wear-out.

One of the performance issues in designing HDD systems is the order of magnitude disparity between random and sequential IO requests. For example, a SATA drive may be able to perform about 100 MB per second for sequential reads but only 0.5 MB per second for small random reads. This makes it difficult to share the drive efficiently across mixed workloads in cloud applications since one random workload could dramatically impact the performance of a sequential workload running on the same set of drives. Write anywhere file systems such as WAFL~\cite{wafl} solve the problem for random versus sequential writes by locating new writes along pre-selected regions of the disk space. However, random reads still retain the large disparity with respect to sequential reads. At the system level, this aspect of drives means that the ideal number of drives and the amount of CPU resources in the controller varies depending on the workload type. 

On the other hand, with Flash, the sequentiality of the IO request stream is less of an issue, though we note that random writes do still affect performance, and so the amount of CPU resources required is better matched to the number of Flash devices. For this reason, Flash is better suited to the mixture of workloads typical of cloud environments, where it is critical to represent different modes over time. In either case -- HDDs or SSDs -- a workload model based on Markov-modulated, net arrival rates and size distributions can capture well the operating modes created by mixing workloads. Irrespective of multi-tenancy cloud workloads, individual applications may have different load levels and operation mixes based on time of day and scheduled background tasks such as backup or mirroring to remote sites. In other words, cloud environments just compound an existing problem. Even during periods of steady input rates from compute servers to shared storage controllers, the underlying storage virtualization software may shape the traffic (for example, to absorb writes quickly and lazily write out to storage later) and this makes the traffic time-varying anyway. This is the post-WAFL level at which we capture our IO traces so that the traffic represents what the physical IO devices will see in their incoming request streams. 

In practice, Flash workloads will differ from traditional HDD workloads, which have been studied extensively in the literature, because the file system will offload some of the accesses from HDDs to Flash. For all file systems, the IO stream to Flash is likely to be more random in nature since this is the workload at which Flash excels compared to HDDs. In the case of the WAFL file system, for example, algorithms detect sequential or patterned read accesses to files and avoid using Flash because these access patterns can be performed well by HDDs, although there are options to bypass the automatic behaviour.

For Flash in particular, we are interested in operation modes that are distinguished by reads only,
writes only, and mixed read/writes. Experimental evidence suggests
that many Flash-based, SSD systems perform less effectively during periods when reads are
mixed with writes. 
\begin{figure}[!h]  \label{vector}
\setlength{\epsfxsize}{0.9 \hsize}
\centerline{\epsfbox{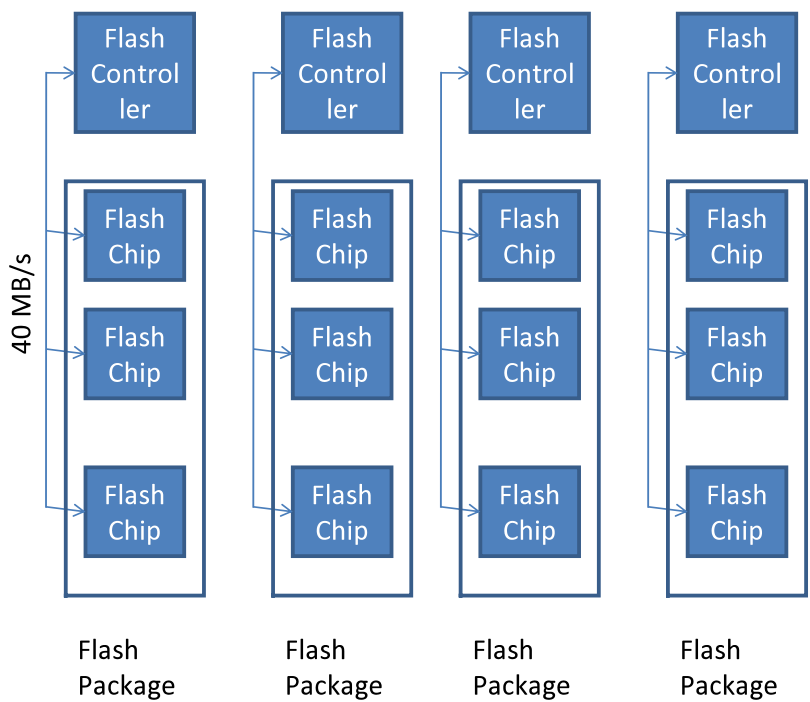}}
\vspace{-1cm}
\caption{Vector of Flash packages with 40 MB/s channels}
\label{vectFIG}
\end{figure}
Indeed our IO trace analysis for the WAFL filesystem (shown in
Figures~\ref{vectFIG} and~\ref{chipFIG}) confirm that read and write rates
vary over time even when the input workload rates are constant. More
generally, we find real system workloads are subject to spikes in
arrival rates, created by data management operations, for
example. Even though these system load perturbations affect the response
times experienced by users, they are often ignored in standard
benchmarks.
\begin{figure}[!h]  \label{chip}
\setlength{\epsfxsize}{1.3 \hsize}
\centerline{\epsfbox{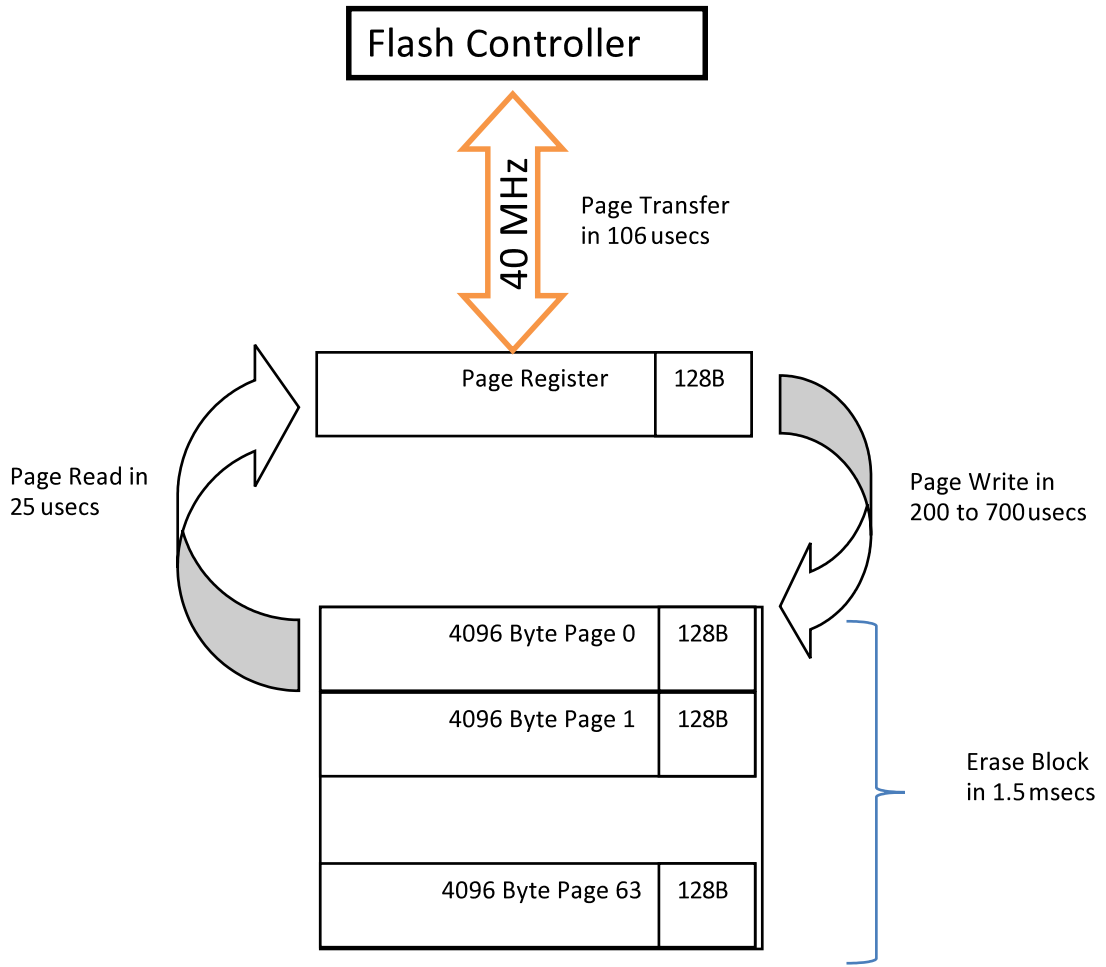}}
\vspace{-2cm}
\caption{Flash chip and controller}
\label{chipFIG}
\end{figure}

\subsection{Why Hidden Markov Models?}
We develop a methodology to characterise the workloads submitted
to an enterprise Flash storage system at various timescales.  Our primary objective is to build portable benchmarks in terms of Markov arrival processes (MAPs), in particular Markov modulated fluid processes (continuous) or Poisson processes (discrete, MMPPs), that can accurately reflect both correlation and burstiness in multi-application workloads that have possibly been pre-scheduled to optimise storage utilisation, availability and access time.  Secondly, at the millisecond timescale, we seek to represent the load at the Flash chip level as input to a previously constructed fluid model that estimates transaction response time distributions~\cite{peva10}.  

In both cases, the workload dynamics is represented by a Markov chain and it is crucial to be able to estimate its generators (or one-step transition probabilities) reliably, to faithfully reflect the key characteristics of the monitored traces.  Such traces are \emph{time series}, which are commonly modelled accurately and in a  parsimonious way by a \emph{Hidden Markov Model} (HMM).  Whilst there are many other approaches to time series analysis, such as the famed Box-Jenkins method and spectral techniques~\cite{boxjenkins,priestly}, a HMM is particularly effective when it is known that the dynamics of the series is strongly influenced by switching between modes, each of which has a particular charcteristic; these modes are represented by the hidden state which evolves as a Markov chain.  Consequently, we focus our attention on identifying a hidden Markov chain that describes the modulation (usually assumed to be Markovian {\it a priori}) of the workload's characteristics at the particular timescale of interest.

\subsection{Related work}
A reliable characterisation of IO workload is essential to the
understanding of the demands placed on a data centre, and particularly important in the construction of
accurate models of performance for which they provide input.  Workload
benchmarks have been constructed routinely over many decades for
profiling, tariffing and system modelling in the form of traces from
typical application mixes, synthetic programs and abstract
mathematical models.
We cite here a selection of such works, ranging from a study
 of a specific parameter's 
 effect on the performance of a particular system to a complete workload characterisation in a generic environment, using a variety of modelling techniques.  
 
Various application behaviours have been analysed to see how critical 
their impact is on data access times in both scientific
and commercial environments; see for example~\cite{reed01,reed02,riska08,wilkes00}.  
The relative impacts of some specifically targeted performance
 parameters are analysed in~\cite{granger04}, where the focus is on the interarrival time, using CART models to express performance predictions as functions of the workload.  
 In~\cite{faloutsos02}, a characterisation of burstiness is extracted and its effect on 
 web and storage workloads is analysed, whilst in~\cite{gomes99}, a structural model suggests 
that disks' IO traces are self-similar, by combining an ON/OFF  source
with a Coxian model.  A more generic, overall workload characterisation is
proposed in~\cite{keaton01a, keaton01b}.
More recently, in~\cite{riska09}, Riska and Riedel analyse enterprise IO disk traces at three time scales -- millisecond, hour and lifetime -- and characterise performance in terms of read/write traffic intensity and disk utilisation.  Burstiness is observed in the traces at each of these three levels, along with long idle periods of the disks.
Again focussing on self-similarity in IO workloads, Wang {\it et. al.} propose a very simple (one parameter) and fast method, called the 
b-model, to generate synthetic, bursty IO traces.
The b-model was validated using real disk and web traces, which matched well with the synthetic
traces in terms of interarrival time and queue length distributions.

In the development of SFS2008 for the CIFS (Common Internet File
System) protocol, the HMM approach was used successfully to model the
desired sequence of CIFS operations from CIFS traces~\cite{sawyer}. The present paper uses
a similar approach (HMMs) but analyses general IO traces and
develops techniques to parameterise MAPs that can be used for
both discrete queueing models and continuous fluid models.

\subsection{Rationale and modelling methodology}
We develop a methodology for representing workloads concisely as a Markov modulated 
process -- a hidden Markov model -- that is suitable both as a portable and flexible benchmark,
with which system administrators can experiment economically, and as
an input to models of storage system performance.  One such performance model is the fluid
model of~\cite{peva10}, which enables the effect of a workload's operation modes on the  
delivered performance of  Flash memory to be assessed.

These workload modelling techniques can be applied to any storage system, but they are especially
appropriate for Flash, in which there is less performance benefit in sequential IO, due to the 
location-independent access time.  
At the same time, the effects of burstiness in each operation mode and correlation are more important to characterise. 
These effects are further amplified by 
file system filters (such as in WAFL), which reschedule the IO streams to optimise
response time and evenly stripe data across a set of storage devices.  In fact, Flash random reads have the same performance as
sequential reads and random user writes are converted into
large sequential or near sequential writes at the device. Flash performance is
sensitive to the mixture of reads and writes, and their sizes, rather than being
dominated by the randomness of the read accesses. Such characteristics
can be well described by an HMM.

The modelling procedure followed below takes trace data monitored on a real system, aggregates it into a discrete set of finite time intervals, called ``bins'',  and then uses cluster analysis to classify the bins into a small set of observation-types.  A Hidden Markov Model is then derived from the sequence of bin-observations by statistical methods, yielding hidden states (also referred to as phases) that represent the input-modes, together with the generators of their Markov chain and the rates at which IO operations are generated in each state.

\subsection{Paper outline}
Following this general introduction on the role and organisation of Flash storage systems, the next section of this paper provides more specific background relevant to our particular Flash workload and HMM model. In section~\ref{model}
we describe our methodology for extracting suitable workload data and the
specific HMM we use, along with an efficient algorithm for estimating its parameters. 
Data taken from a Flash
storage system is analysed and aggregated in section~\ref{tests}, leading to an HMM that is validated against the raw data.   This HMM is combined with a fluid model of Flash performance in section~\ref{FlashModel} and preliminary results on response time distributions are displayed.
Section~\ref{concl} summarises this work and outlines plans for further investigation.  

\section{Flash workload model}
\label{BackMotiv}
In this section, three use cases are identified for our proposed workload modelling methodology and HMMs are introduced as the means for constructing synthetic benchmarks.

\subsection{Three use cases} 
Our workload characterisation methodology provides a portable Markov modulated workload model, for which there are three distinct use cases:
\begin{enumerate}
\item Generating workload traces for live systems from which typical performance measurements are required; 
\item Generating similar, perhaps more abstract, traces for system simulation; 
\item Providing input parameters for analytical performance models.  
\end{enumerate}

The first two of these are similar in functionality but there will usually be a difference in the level of detail in the traces since a model is inherently more abstract than the real system it represents.  
In the present work, we apply the output of the derived HMM to the second and third use cases.  
The HMM approach represents trace-level operation sequences in a compact form and we demonstrate that the original traces and the HMM-generated traces are comparable as far as key performance metrics are concerned.  
In the second use case, raw and HMM-modelled traces are generated as input to a hardware simulator.  The faithfulness of the HMM is assessed by comparing the mean simulated queueing times of requests in the raw and HMM-generated traces\footnote{Stronger validation might compare histograms of the queueing times predicted by the simulation model, for which a more detailed workload abstraction would likely be needed; e.g. the traces' time-stamps, which we aggregate out into bins.  This would require a much more complex HMM based on a continuous time hidden Markov chain}.  
The third use case supplies input parameters for a Markov modulated fluid model of Flash storage systems, as in~\cite{peva10} for example.  The outputs of this model and of a hardware-customised simulator were compared by way of mutual validation of these two system models.

\subsection{Hidden Markov Models} 

A Hidden Markov Model (HMM) is a bivariate Markov chain, which we will denote by
$\{ (C_k,S_k)\} _{k\ge0}$, where $k$ is an integer index. The related Markov chain $\{ C_k\}$
is \emph{hidden} in the sense that its states are not directly
observable. What is available to the observer is another stochastic
process   $ \{ S_{k} \} _{k\ge0}$, linked to the hidden Markov chain in that
$C_k$ governs the distribution of the corresponding $S_k$. All
statistical inference, even on the Markov chain itself, has to be done
in terms of $\{S_k\}$ only, as $\{C_k\}$ is not observed. A further
assumption is that $C_k$ must be the only variable of the Markov chain
that affects the $S_k$ distribution.

HMMs and their generalisations are used nowadays in
many different areas, essentially to provide parsimonious
descriptions for time series that can be very long (a billion points). 
They are especially known for their application in temporal pattern recognition such as speech, handwriting, gestures, musical scores and bioinformatics~\cite{rabiner,gesture,music}.
A HMM can provide the input to the simulation of a complex system in a faithful, concise way and its suitability for modelling Flash workloads of the type discussed above is clear.

There are three main issues in the construction of a HMM. First, given the
parameters of the model, compute the probabilities of a particular
sequence of observations, and of the corresponding hidden state
values. These probabilities are given by the so-called forward-backward equations~\cite{skinnybook}.  
Secondly, given a sequence of observations, or set of
such sequences, find the most likely set of model parameters. This may be solved, using \emph{a priori}
statistical inference, by the Baum-Welch Algorithm, 
which itself uses the forward-backward algorithm.  Thirdly, find the most likely sequence of
hidden states that could have generated a given sequence of
observations. This is done using \emph{a posteriori} statistical
inference in the Viterbi Algorithm~\cite{skinnybook}. 

\begin{definition} \label{HMMdef}
Suppose that $\{C_t\}_{t=0,1,\ldots}$ is a Markov chain with state space $ {\cal S} = \{1,\ldots,r\}$,  initial distribution $\nu_c$ $(c \in {\cal S})$ and transition matrix $Q=(q_{cc'})_{c,c' \in {\cal S}}$, and $\{S_t\}_{t=0, 1, \ldots}$ is a stochastic process taking values in ${\cal J} =\{1,  \ldots, m \}$.   The bivariate stochastic process $\{(C_t, S_t)\}_{t=0,1,\ldots}$ is said to be a \emph{Hidden Markov Model} if it is a Markov chain with transition probabilities 
\begin{eqnarray*}
P(C_t=c_t, S_t=s_t | C_{t-1}=c_{t-1}, S_{t-1}=s_{t-1})=\\
P(C_t=c_t, S_t=s_t | C_{t-1}=c_{t-1})=q_{c_{t-1}c_t}g_{c_ts_t}
\end{eqnarray*}
where $G=(g_{cs})_{c \in {\cal S}, s \in {\cal J}}$ is a stochastic matrix.
\end{definition}
A very useful property is that, conditionally on $\{C_t\}_{t=0,1,\ldots}$, $\{S_t\}_{t=0,1,\ldots}$ are independent, i.e.
$$P(S_0^n=s_0^n | C_0^n=c_0^n) = \prod_{i=0}^n P(S_i=s_i | C_i=c_i)$$
where $x_i^j$ denotes the sequence $(x_i, x_{i+1}, \ldots, x_j)$.
This result is used extensively in the derivation of the enhanced Baum-Welch and Viterbi algorithms, described below, which we use in our analysis.

\section{From traces to workload model}
\label{model}
\subsection{Levels of offered workload}

\begin{figure}[!h]  \label{stack}
\setlength{\epsfxsize}{0.9 \hsize}
\centerline{\epsfbox{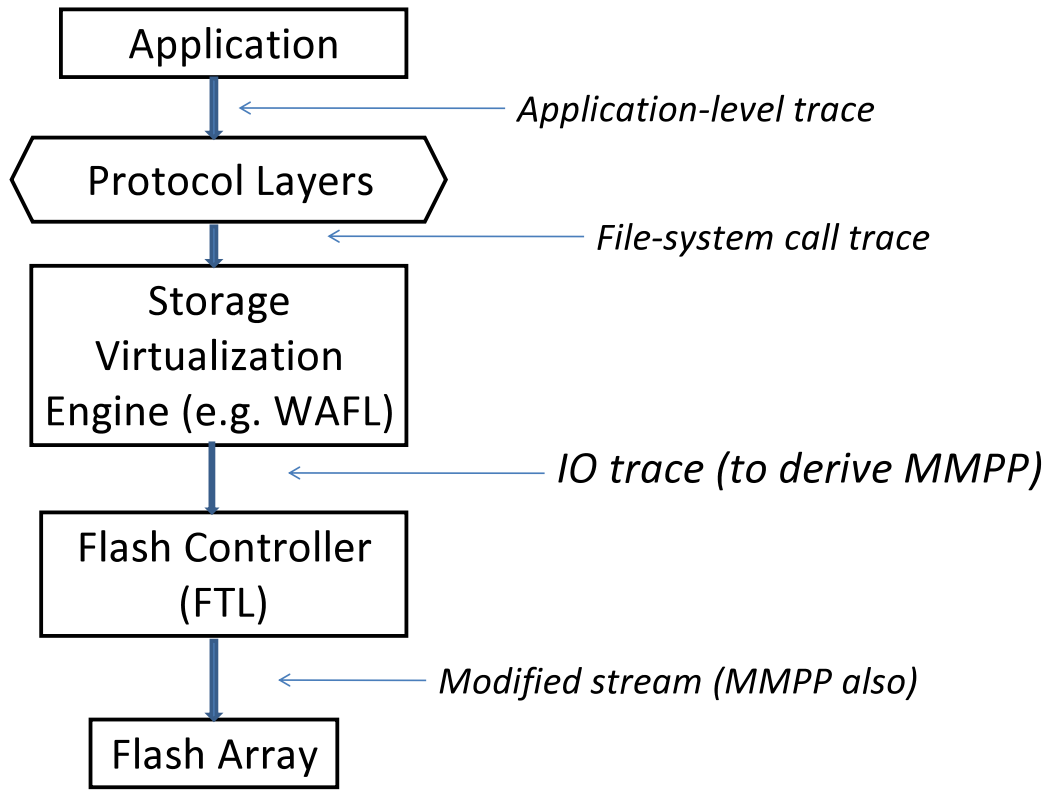}}
\vspace{-0.7cm}
\caption{\label{stackFIG} Flash storage stack}
\end{figure}

Workloads typically go through a number of transformations as they
traverse the storage controller stack (Figure~\ref{stackFIG}). For example, the read sequences
arriving at the input ports are filtered by caching, increased by
file system overhead/prefetch etc. and eventually emerge at the IO
devices as a physical read request stream. Writes
and metadata operations often have more complex transformations across
the stack. In the case of Flash, the workload will pass through a 
storage virtualisation processor and a
Flash Translation Layer (FTL) that can introduce additional traffic
(write amplification) based on the specific Flash design choices
implemented. In storage systems, we can collect traces at various
layers of the stack; we collected traces after storage virtualisation 
but before the FTL processing in the Flash packages --
this was the lowest level at which we could practically collect
detailed traces.  Even though the details of the Flash packages' design
choices are often unknown externally, we can represent known FTL
design features by appropriately increasing traffic rates in a MAP workload model;
for example, to account for expected write amplification on the given
workload with a known amount of free space.

\subsection{Discrete ``binned'' time series}
For Flash, sequentiality of reads does not affect performance, so
we can simply count the number of read blocks (per bin), rather than considering
the length of sequential read chains. Since the WAFL file
system converts random user writes into sequential writes we can also
consider just the counts of the write blocks. More generally, to allow for deferring of writes, we might also need
to represent the write block run length (or write chain) distribution,
or at least its average, although this could be accounted for in our model by burstiness, 
represented by the time spent in a hidden state associated with writes.  This issue is not addressed specifically in this paper, however.  
For example, we do not predict individual, time-stamped workloads in which the precise impact of random accesses can be assessed.   
Modelling at this level of detail is close to emulation and punitive computationally.
We are concerned with how the workload offered to a storage system varies (in intensity and type) over time.  
The level of abstraction we consider is bursts of IO activity characterized by predominant levels of operation type, for example reads, writes or mixed, together with their correlation over time.  Whilst particularly well suited to Flash workloads for the reasons stated above, the method is equally applicable to other types of file system from which time series are constructed by appropriate choices of bin size and clustering algorithm;  see for example section~\ref{user-workload}.  

The raw trace data consists of time-stamped records containing information about the IO operation that arrived at the given time.  A more manageable and understandable representation is provided in summary form by partitioning the time line into a series of uniform ``bins'', i.e. contiguous intervals of some chosen, constant size.  For each bin, we compute from the raw, time-stamped trace, a time series consisting of summary information: at least the number of IO operations of each type arriving during the time interval associated with the bin as well as other workload data such as the sizes of each IO.  Such time series are amenable to standard analysis techniques, involving the autocorrelation function and the power spectrum.  The width of the bins is chosen according to the timescale required for the modelling exercise.  For constructing a portable workload model, this will depend on the level of detail required and intended purpose of that model.  

For example, for long term planning, the evolution of the numbers of IO operations per hour or even days might be appropriate, suggesting bins with a similar width.  
At the other extreme, to parameterise a detailed performance model at
the Flash chip level, the raw data may be  transformed into a discrete
time series by characterising it in 5ms or 1ms, bins.  Each one accumulates
a vector of counts: one count for each read of size $N$ units and one
for each write of size $M$ units, for all possible sizes $N,M$.  This
binned trace may be further summarized into a vector of pairs, (number
of read blocks, number of write blocks) or $(R,W)$, in each interval.  We call this the
summary trace, which is used to derive the Markov modulated fluid arrival process describing the input
to the fluid model in~\cite{peva10}.
 
 A bin size of five milliseconds was found to be suitable for the post-WAFL traces, representing the workload seen by the Flash chips, whereas for the pure user-generated workload considered in section~\ref{user-workload}, we found bin sizes of no less than one second to be best.  Too small a time
interval creates excessive noise and tends to produce many intervals that are empty, 
or with very small IO counts, whereas 
too large an interval tends to miss some characteristics of
the low-level sequence of operations, for example mode transitions.  
The chosen raw binned traces are used to construct more abstract model inputs and to glean more detailed information about the underlying user workloads and storage management systems (e.g. WAFL) that transform them. This is done by applying a standard clustering algorithm, so that points `nearby' in some Euclidean space are assigned to the same subset out of a small partition; the number of subsets is specified by the user.  The points in each subset, or cluster, are used to create a mapping from the binned data point $(R,W)$ to an \emph{observation value}, e.g. to the pair (centroid, standard deviation) of the points in the cluster. The set of observation values needs to  be relatively small (e.g. between about 10 and 20) so that they can be interpreted physically and the time required to compute the HMM is manageable. The backward mapping from observation value to data point, usually probabilistic, is also important for understanding the characteristics of each observation value and for generating synthetic workloads comprising typical sequences of data-points.

\subsection{Construction of a HMM from trace data}  \label{constructionofHMM}

The joint probability function of $C_0, S_0, \ldots ,C_n, S_n$ can be
written as:
\begin{eqnarray*}
J_{\nu, n}(c_0^n, s_0^n) &=& P(C_0=c_0,S_0=s_0,\ldots,C_n=c_n,S_n=s_n) \\
&=& P(C_0=c_0,S_0=s_0) \times \\
&& \quad P(C_1=c_1,S_1=s_1 | C_0=c_0) \ldots P(C_n = c_n, S_n = s_n | C_{n-1} = c_{n-1}) \\
&=& \nu_{c_0}g_{c_0,s_0} \prod_{i=1}^n q_{c_{i-1},c_i}g_{c_i,s_i}
\end{eqnarray*}
This is the full likelihood function pertaining to both unobserved and observed random variables.
The likelihood of the observations $s_0, s_1, \ldots, s_n$ is defined as
$$L_{\nu, n}(s_0^n) = L_{\nu, n}(s_0, s_1, \ldots, s_n) = P(S_0=s_0, \ldots, S_n=s_n)$$
so that 
\begin{eqnarray*}
L_{\nu, n}(s_0^n) &=& \sum_{c_0, \ldots, c_n} J_{\nu, n} (c_0^n,  s_0^n)\\
&=& \sum_{c_0, \ldots, c_n} \nu_{c_0}g_{c_0,s_0} \prod_{i=1}^n q_{c_{i-1},c_i}g_{c_i,s_i}
\end{eqnarray*}

The binned, summary workload trace is used to estimate the parameters of a HMM, i.e. the transition probabilities of the Markov chain followed by the hidden state, the probabilities of seeing each possible observation (defined in terms of the characteristics of a bin) in each of the hidden states, and the initial probability function of the hidden state.  This is done using the {\em normalized Baum-Welch algorithm} -- a modification of the original Baum-Welch algorithm that uses scaling to avoid the underflow that is inevitable when considering raw likelihood functions over very long sequences.  We first define the following smoothing functions:

\begin{definition}
For positive integers $k \leq \ell \leq n$, let
\begin{eqnarray*}
\phi_{k:\ell | n}(c_k^\ell) = P(C_k=c_k, \ldots, C_\ell = c_\ell \mid S_0=s_0, \ldots, S_n=s_n)
\end{eqnarray*}
We use the abbreviation $\phi_{k:k|n}(j,j)=\phi_{k|n}(j)$. 
\end{definition}
Clearly the above likelihood function can be expressed in terms of these smoothing functions, and it is special cases of $\phi_{k:\ell | n}$ that are computed by the aforementioned forward-backward equations.  The parameters of the sought after HMM are now given by the following result:

\begin{proposition} (Normalized Baum-Welch Algorithm)\\
Optimal re-estimates for the HMM parameters are, for $1 \leq j,k \leq r, 1 \leq s,s_i \leq m$,
\begin{eqnarray*}
\hat{\nu_j} &=& \frac{  \phi_{0 \mid n}(j)} {\sum_{\ell=1}^r  \phi_{0 \mid n}(\ell)}\\
\hat g_{js}&=& \frac{ \sum_{i=0}^{n} \delta_{s_i,s}  \phi_{i \mid n}(j)}{\sum_{i=0}^{n}  \phi_{i \mid n}(j)} \\
 \hat q_{jk}&=& \frac{\sum_{i=0}^{n-1}\phi_{i : i+1 \mid n}(j, k)} {\sum_{i=0}^{n-1} \phi_{i \mid n}(j)}
\end{eqnarray*}
where $\delta$ is the Kronecker delta defined by $\delta_{st}=0$ if $s \neq t$ and $\delta_{ss}=1,~(1 \leq s,t \leq m$), $r$ is the number of hidden states, $m$ is the number of observation values, and $n$ is the length of the given time series.  
\end{proposition}
\medskip
The proof, sketched in the Appendix, uses the Expectation Maximization (EM) principle and is developed in terms of the smoothing function, $\phi_{k:\ell | n}$.  It should be noted that it is these smoothing functions that are normalized in order to ensure numerical stability in the Baum-Welch algorithm, and others~\cite{fatbook}.  

Each iteration involves matrix-multiplications to apply the forward-backward algorithm and to update the Baum-Welch estimates, giving a computation time of $\Theta(n r^2 + n r m)$ per iteration.   Since, in practical examples, there will be more observation values than hidden states, $m>r$ and so each iteration has time complexity  $\Theta(n r m)$. The asymptotic time complexity of the complete Baum-Welch algorithm is therefore  $\Theta(h n r m)$, where $h$ is the number of iterations needed for convergence, and the storage requirement is easily seen to be $\Theta(r n)$, the space required to hold the smoothing functions.

Next, the {\em Viterbi algorithm}, introduced below, is applied to the
binned trace to estimate the mean durations of the various modes of operation (e.g. read or write), as represented by the ``hidden'' state.  Let $m_i$ be the average number of consecutive observations instants at which the hidden state $i$ is the same, according to the Viterbi output trace.  The mean holding time of state $i$ is then estimated as the product of $m_i$ and the bin-width, $5$ms.  Together with the state transition probabilities, these mean holding times determine the generators of the resulting MAP that is used to parameterise the workload input to the Flash fluid model~\cite{peva10}.  The generator (or instantaneous rate) $a_{ij}$ is estimated as the transition probability from state $i$ to state $j\neq i, q_{ij}$ (obtained from the Baum-Welch algorithm), divided by $0.005 \, m_i$, estimated by Viterbi.  Note that the Viterbi algorithm is not needed to generate discrete, binned, synthetic workload traces, but only to construct a continuous time workload process.  In fact, a continuous time Markov chain (CTMC) could be obtained directly from the one-step self-transition probabilities $q_i=-q_{ii}$,  estimated for each state $i$: the mean number of consecutive bins in which the state remains at $i$ would be $1/(1-q_i)$ in the derived Markov chain, this number being a geometric random variable.  However, this calculation is rather unstable since the denominator is very small for even moderately long state holding times and so is highly sensitive to the precision of the estimate of $q_i$.  We have found the Viterbi method to be much more accurate.  

Finally, access-size probability distributions for each type may be 
estimated empirically by constructing histograms from the sizes of
every IO submitted during each of these operation types, as identified by Viterbi.  

As with the Baum-Welch algorithm, an enhanced version of the standard
Viterbi algorithm is required, using log-likelihoods instead of plain
likelihoods to avoid underflow. For ease of reading, henceforth
the observations 
$S_0, \ldots, S_n$ are assumed to be fixed. We write
\begin{eqnarray*}
\phi_{0:k+1 \mid k+1}(c_0,\ldots,c_{k+1}) &=& P(C_0^{k+1}=c_0^{k+1} \mid S_0^{k+1}=s_0^{k+1}) =
\frac{J_{\nu, k+1} (c_0^{k+1}, s_0^{k+1})}{L_{\nu, k+1} (s_0^{k+1})} \\
&=& \frac{L_{\nu,k}}{L_{\nu, k+1}}~\phi_{0:k \mid k}(c_0,\ldots,c_{k}) q_{c_{k} c_{k+1}} g_{c_{k+1} s_{k+1}}
\end{eqnarray*}
Denoting $l_k=\log L_{\nu, k}$, we have
\begin{eqnarray} \label{logphi}
\lefteqn{ \log \phi_{0:k+1 \mid k+1} (c_0, \ldots, c_{k+1}) = } \hspace{1cm} \nonumber\\
&&  l_k - l_{k+1} +  \log (\phi_{0:k \mid k} (c_0, \ldots, c_{k}))+ \log (q_{c_{k} c_{k+1}} g_{c_{k+1} s_{k+1}}) 
\end{eqnarray}
so that
\begin{eqnarray}
\lefteqn{\log \phi_{0:k+1 \mid k+1} (c_0, \ldots, c_{k+1}) + l_{k+1}=} \hspace{1cm} \nonumber \\
&& l_k  + \log (\phi_{0:k \mid k} (c_0, \ldots, c_{k}))+ \log (q_{c_{k} c_{k+1}}) +\log( g_{c_{k+1} s_{k+1}})
\end{eqnarray}
We now define
\begin{eqnarray} \label{m}
m_k(i) = \max_{c_0, \ldots, c_{k-1}} \log (\phi_{0:k \mid k} (c_0, \ldots, c_{k-1}, i)) + l_k
\end{eqnarray}
Up to the constant term $l_k$, this is the maximum \emph{a posteriori} log probability of sequences  up to time $k$ that end with state~$i$.\\
For $k>0$, let $b_k(i)$ be the value of $c_{k-1}$ for which the maximum is achieved in equation \ref{m}. That is, $b_k(i)$ is the second final state in an optimal state sequence (up to time $k$) ending with state $i$. Now, using equation \ref{logphi}, we have
\begin{equation} \label{mk+1}
m_{k+1}(j) = \max_i (m_k(i) + q_{ij}) + \log (g_{j s_{k+1}})
\end{equation}
and $b_{k+1}(j)$ is the $i$ for which the maximum is achieved.
For $k=0$, we have $m_0(i) = \log (\phi_{0 \mid 0}(i) L_{\nu, 0}) = \log (P(C_0=i \mid S_0=s_0)P(S_0=s_0)) = \log (\nu (i) g_{i s_0})$

Starting with $m_0(i) = \log (\nu(i) g_{i s_0})$ and running this recursion for $k=1, \ldots,{n-1}$, we obtain the final $m_n(j)$. By maximizing $m_n(j)$ w.r.t $j$ we can identify the final state $\hat c_n$ of the optimal sequence of states. So  $\hat c_n$ is the value of $j$ that maximises $m_n(j)$. Now, the optimal sequence can be found by running the backward recursion
$$\hat c_k = b_{k+1}(\hat c_{k+1}), ~~~~~~~~~k=n-1, \ldots,0$$
that is,  $\hat c_k$ is the value of the second-last state for which the maximum of $m_{k+1} (\hat c_{k+1})$ is attained.  This is summarized in Viterbi's algorithm:
\begin{algorithm}{{(Viterbi Algorithm)}}
\begin{enumerate}
\item Forward recursion (for optimal conditional probabilities): Let
$$m_0(i)=\log (\nu (i) g_{i s_0})$$
Then for $k=0,1,\ldots,n-1$ and for all states $j$, compute 
$$m_{k+1}(j) = \max_i (m_k(i) + q_{ij}) + \log (g_{j s_{k+1}})$$
\item Backward recursion (for optimal sequence): Let $\hat c_k$ be the state $j$ for which $m_n(j)$ is maximal. Then for  $k=n-1,n-2,\ldots,0$, let $\hat c_k$ be the state $i$ for which the maximum is attained  for $m_{k+1}(j)$ in the forward step, with $j=\hat c_{k+1}$. That is, $\hat c_k=b_{k+1}(\hat c_{k+1})$.
\end{enumerate}
\end{algorithm}

\begin{figure}[!h]
\setlength{\epsfxsize} {0.9 \hsize}
\centerline{\epsfbox{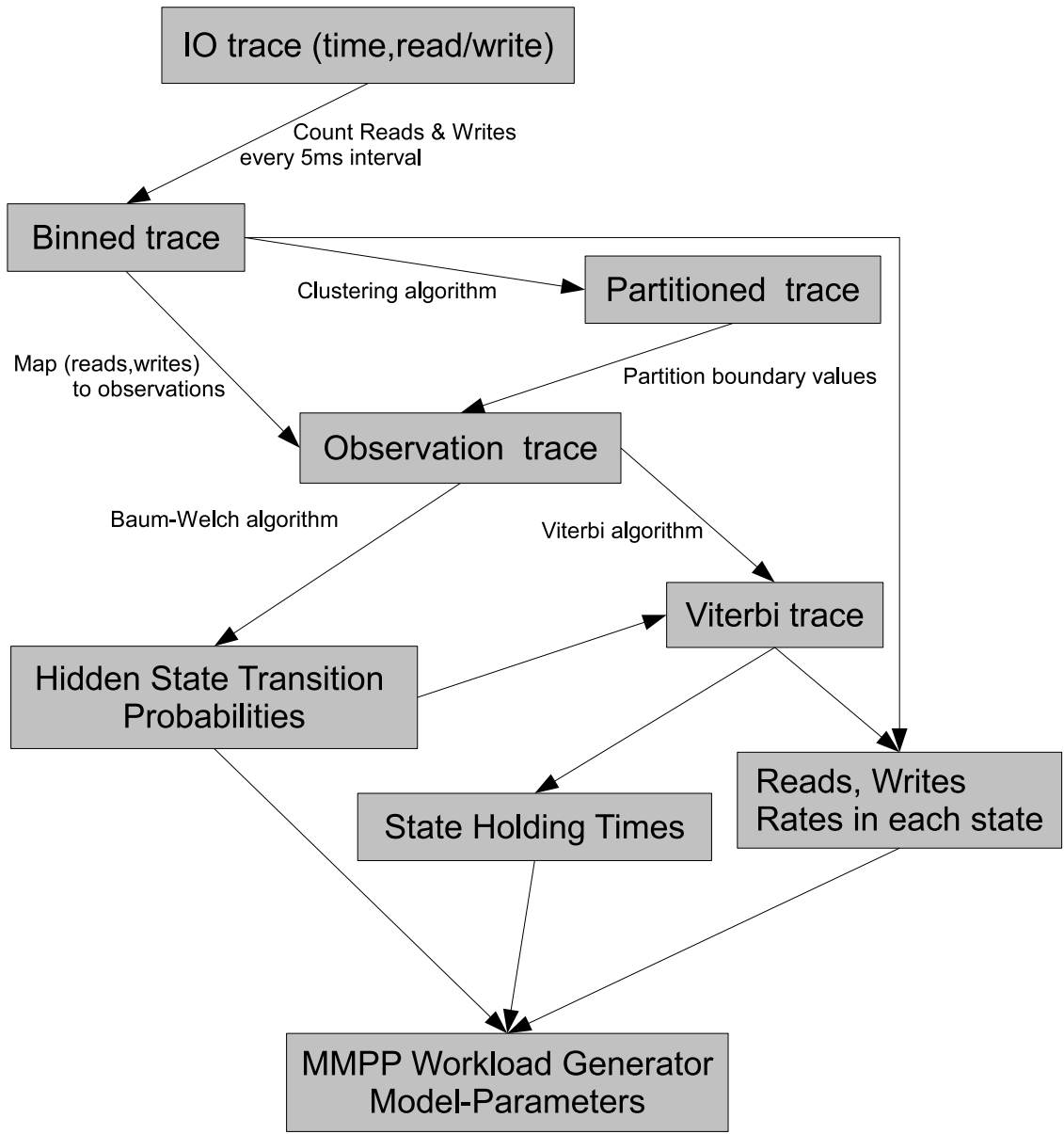}}
\vspace{-3cm}
\caption{IO trace analysis methodology: from trace to MAP parameters}
\label{fig4}
\end{figure}

The complete workload modelling methodology is summarized in Figure~\ref{fig4}, which shows the mapping of the raw trace-data into binned form, through clustering into a trace of observation values, which is processed by the Baum-Welch and Viterbi algorithms to produce MAP (here, Markov modulated fluid process) generators.
An HMM for a particular workload derived using this methodology can represent subtle interactions that are often lost in simple workload models. For example, reads and writes are often treated independently but in our workload model we can represent the natural feedback that lowers read rates during periods of high write rates, and vice-versa. Although the HMM generates traffic that does not depend on the state of the system, it does account for the dynamic feedback through its choice of parameters associated with the hidden states.

\section{Flash chip workload model and its validation}
\label{tests}
\subsection{HMM workload model} \label{HMMwlmodel}
We used two IO traces derived from the SPC-1
 benchmark. In the first run, this benchmark generated about 61\% writes and 39\%
 reads, and also  included a mixture of random reads and writes as well as
 sequential reads and writes.  Being write-dominated, we call this the \emph{update-mix} run.  The second trace was read-dominated, intended to represent on-line transaction processing workloads, with about 67\% reads, which we call the \emph{oltp-mix} run.
Each case produced an IO trace of time-stamped read or write IO operations,
 together with the operation size (in 4KB units). From this we used 5 ms bins
 to create a ``binned trace'' consisting of a pair of values: number of
 read blocks and number of write blocks.  This trace was further
 reduced by applying a standard clustering algorithm (available in Mathematica Version 7~\cite{wolfram}).  
Using two partitions for reads and four partitions for writes, we defined eight observation values by the corresponding centroids and mapped each of the read-write pairs to one of these eight. 
The vectors of eight centroids that defines the set of observation values for each case (respectively \emph{update-mix} and \emph{oltp-mix})  are the following:
 
$$ {\footnotesize 
\left(
\begin{array}{ll}
 5.7 & 0.28 \\
 4.45 & 31.3 \\
 5.11 & 82.1 \\
 4.49 & 183.0 \\
 23.7 & 0.217 \\
 24.5 & 31.7 \\
 27.3 & 81.0 \\
 25.8 & 165.8
\end{array}
\right) }
\;\;\;\mbox{and}\;\;\;
{\footnotesize \left(
\begin{array}{ll}
37.3  & 0.18 \\
32.3 & 22.4 \\
32.5 & 36.1 \\
33.1& 84.6\\
139.5 & 0.14 \\
126.1 & 22.2 \\
140.1 & 35.6 \\
141.4 & 81.8
\end{array}
\right)
} $$
Observation values of 1 to 4 represent low reads and increasing writes. Similarly, values 5 to 8 represent high reads and increasing writes.

The sequence of observation values makes up the observation trace that is fed into the Baum-Welch algorithm, with three hidden states chosen.  The choice of the number of states was by trial and Occam's razor.  Three states was found to be superior to two states by essentially identifying more modes of operation; for example, in terms of predominantly small or large reads and writes.  However, using four hidden states led to a pair of states with very similar characteristics in their $G$-matrix rows and such that their transitions (in the matrix $Q$) were either almost the same as well or else oscillated between each other with very high probability.  Such a pair is better replaced by a single state.  The transition probability matrix $Q$ and the observation value probabilities $G$ produced by the Baum-Welch algorithm are used in the Viterbi algorithm to compute the expected hidden state for each trace point. This ``Viterbi trace'' of hidden state values is used to estimate the mean state holding times in the Markov modulated fluid input process of section~\ref{FlashModel}; see section~\ref{constructionofHMM}.  Further, the read and write rates for each state can then be approximated by looking back at the binned trace. Altogether, this gives sufficient information to parameterize the MAP workload generator model.

From the IO trace we used, the Baum-Welch algorithm produced the following transition probability matrices for the HMM  (to 4 decimal places), again for the \emph{update-mix} trace (left) and the \emph{oltp-mix} trace (right): 
$$ {\footnotesize 
Q = \left(
\begin{array}{ccc}
 0.9972 & 0.0022 & 0.0006 \\
 0.0005 & 0.9965 & 0.0030 \\
 0.0021 & 0.0005 & 0.9974
 \end{array}
\right) }
\;\;\;\mbox{and}\;\;\;
{\footnotesize 
\left(
\begin{array}{ccc}
 0.9917 & 0.0005 & 0.0077 \\
 0.0248 & 0.9726 & 0.0026 \\
 0.0022 & 0.0136 & 0.9841
 \end{array}
\right) 
} $$
The diagonal values relate to the ``run length'' of each state in that they are the probabilities that the state remains the same in the next time slot, i.e. bin.  The expected run length is $1/(1-d)$ for a state with diagonal entry $d$.  This is of the order of $100$ bin-widths here, representing a time of around half a second, which is quite reasonable.

The probabilities of seeing each of the observation values in each of the hidden states were computed, for the two traces respectively, as (to 3 decimal places), 
$$ {\footnotesize 
G =\left(
\begin{array}{cccccccc}
 0.634 & 0.0 & 0.0 & 0.0 & 0.366 & 0.0 & 0.0 & 0.0 \\
 0.076 & 0.118 & 0.237 & 0.112 & 0.068 & 0.115 & 0.199 & 0.076 \\
 0.241 & 0.393 & 0.002 & 0.0 & 0.158 & 0.205 & 0.0 & 0.0 
\end{array}
\right) 
} $$
$$ \mbox{~~~and~} {\footnotesize 
\left(
\begin{array}{cccccccc}
 0.153 & 0.064 & 0.252 & 0.315 & 0.060 & 015 & 073 & 067 \\
 0.786 & 0.0 & 0.0 & 0.0 & 0.214 & 0.0 & 0.0 & 0.0 \\
 0.507 & 0.048 & 0.207 & 0.0 & 0.168 & 0.018 & 0.52 & 0.0 
\end{array}
\right) 
} $$
Finally, the initial hidden state probability vectors were estimated as (to 3 decimal places)
$(0.0, 0.592, 0.408)$ and $(0.036, 0.446, 0.518)$.

In a real Flash storage system, the filesystem performs consistency points (CPs or checkpoints) that write data to permanent storage in a consistent way.  We see from the output of the Viterbi algorithm (not listed here) that the HMM-generated traces exhibit oscillating behavior in that the writes appear in relatively long bursts, interspered with periods of predominantly read-behavior.  In the HMM, state 1 issues mainly reads (see the matrix $G$) and these periods can be thought of as ``outside of CP". States 2 and 3 in the \emph{update-mix} trace represent different levels of write activity, with state 2 delivering higher write rates, and have relatively long runs in the associated HMM Viterbi-trace. This matches the long periods during which CPs are occurring.  A similar interpretation applies to the \emph{oltp-mix} trace.  

\subsection{Validation of the HMM}
We validated the HMMs using Monte Carlo simulation to generate from them traces of the numbers of 
read and write blocks in a sequence of 5ms bins of specified length -- chosen to be the same as for the raw traces.  
The parameters of the HMMs were set to the estimates derived from the corresponding raw traces.  
The key metrics of interest are the means
 and standard deviations of the numbers of read and write blocks in each bin.   
 We first compared the raw (original binned) trace with HMM-generated traces in terms of the observation values, i.e. cluster numbers.  
 For the raw trace, the cluster numbers were those to which the data-point in each bin belonged, 
 i.e. the trace was precisely that used as input to the Baum-Welch algorithm.  
The mean and standard deviation of the read and write block-counts per bin over the whole trace were computed using the centroid values for the 
cluster assigned to each bin in each trace.  
These traces were in very good agreement according to the key metrics but, clearly, clustering the reads and writes in this way loses significant information.  The results are not shown here.  

Instead, for each bin in the HMM trace, we generated a pair of read and write block-counts by assuming a bivariate Normal distribution (truncated at zero) with means and covariance matrix given by the centroid, standard deviations and read-write block-count covariance (not shown here) of the associated cluster.

 \subsubsection{Means and standard deviations}
After a 5ms-binning, the original binned traces (for \emph{update-mix} and \emph{oltp-mix} workloads) have the summary
performance metrics (mean counts and their standard deviations per bin) shown in Table~\ref{table1}.  
These data are compared with the corresponding HMM-traces, 
derived from the Baum-Welch algorithm from which the read-write block-count pairs were generated as described above.    
Confidence bands at the $96\%$ level were computed for the (Monte Carlo) HMM trace estimates using the standard batch-means method; ten HMM traces were used for each of the two workload mixes.  These are also shown in Table~\ref{table1}.

 \begin{table} 
 {\footnotesize 
\begin{tabular}{|c||c|c|c|c|c|c|c|c|} \hline
 & \multicolumn{4}{c|}{Reads/bin} & \multicolumn{4}{c|}{Writes/bin} \\ \hline 
 Workload & \multicolumn{2}{c|}{Raw~~~}  & \multicolumn{2}{c|}{HMM~~~} & \multicolumn{2}{c|}{Raw~~~} & \multicolumn{2}{c|}{HMM~~~}  \\ \hline
type & Mean & Std Dev  & Mean & Std Dev & Mean & Std Dev  & Mean & Std dev \\
\hline \hline 
\emph{update-mix} & 12.71 & 12.15 & $\begin{array}{c} 12.97 \\ 12.59 \end{array}$ & $\begin{array}{c} 12.11 \\ 11.77 \end{array}$  & 27.98 & 44.82 & $\begin{array}{c} 32.53 \\ 23.12 \end{array}$ & $\begin{array}{c} 49.40 \\ 40.73 \end{array}$  \\
\hline 
\emph{oltp-mix}  & 57.96 & 53.55 & $\begin{array}{c} 58.84 \\ 57.89 \end{array}$ & $\begin{array}{c} 53.43 \\ 52.59 \end{array}$  & 28.12 & 34.71 & $\begin{array}{c} 29.30 \\ 26.17 \end{array}$ & $\begin{array}{c} 34.94 \\ 33.57 \end{array}$ \\
\hline
\end{tabular} }
\centering{
\caption{ \label{table1} Bin means and standard deviations for raw and HMM traces, with 96\% confidence bands. }
}
\end{table}
The bin-means of the HMM trace match those of the raw trace very well.  More pleasingly, the standard deviations are also in good agreement, mostly underestimating the raw values a little.  Only in two cases do the raw standard deviations lie slightly outside the confidence interval: for the standard deviations of the read block-counts.  For the \emph{update-mix} and \emph{oltp-mix} workload respectively, the raw estimate, based on measurement, exceeded the upper  $96\%$ confidence level by only about $0.33\%$ and $0.22\%$.

\subsubsection{Correlation}
Analytical models of storage resources often use relatively simple
models for arrival streams, such as just the mean and standard deviation of the
inter-arrival times for IO requests. However, in practice we find from IO
traces that these IO streams are rarely independent,
especially when they are filtered through a file system that optimizes
for performance. The challenge is to represent the correlations within and 
among streams with a small set of parameters that can be readily
derived from trace data.  This certainly applies to a Flash
system using WAFL to reschedule its incoming IO streams in order
to optimise response time.
We compared the autocorrelation functions (ACF) of the raw, unclustered traces (input to the Baum-Welch algorithm) with those of corresponding HMM-generated traces.  

\begin{figure}[!h]
\centerline{
{\setlength{\epsfxsize}{0.5 \hsize}
\epsfbox{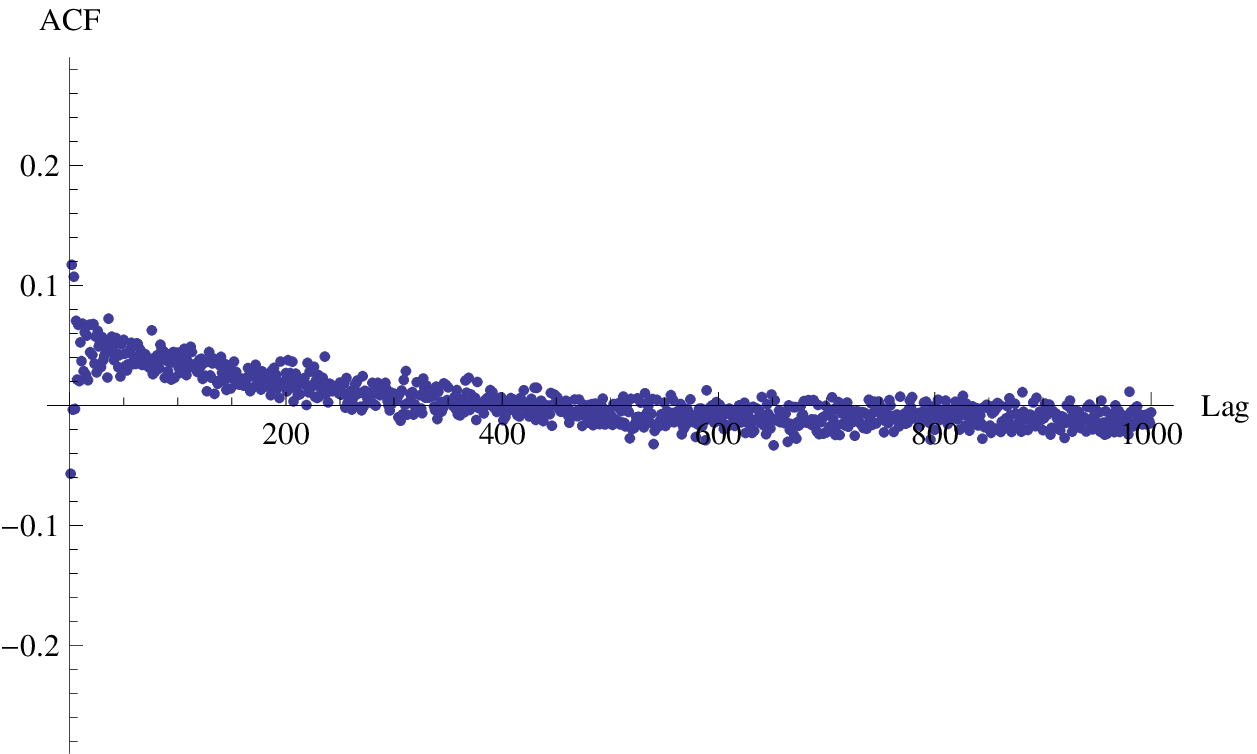}}
\hspace{-0.1in}
{\setlength{\epsfxsize}{0.5 \hsize}
\epsfbox{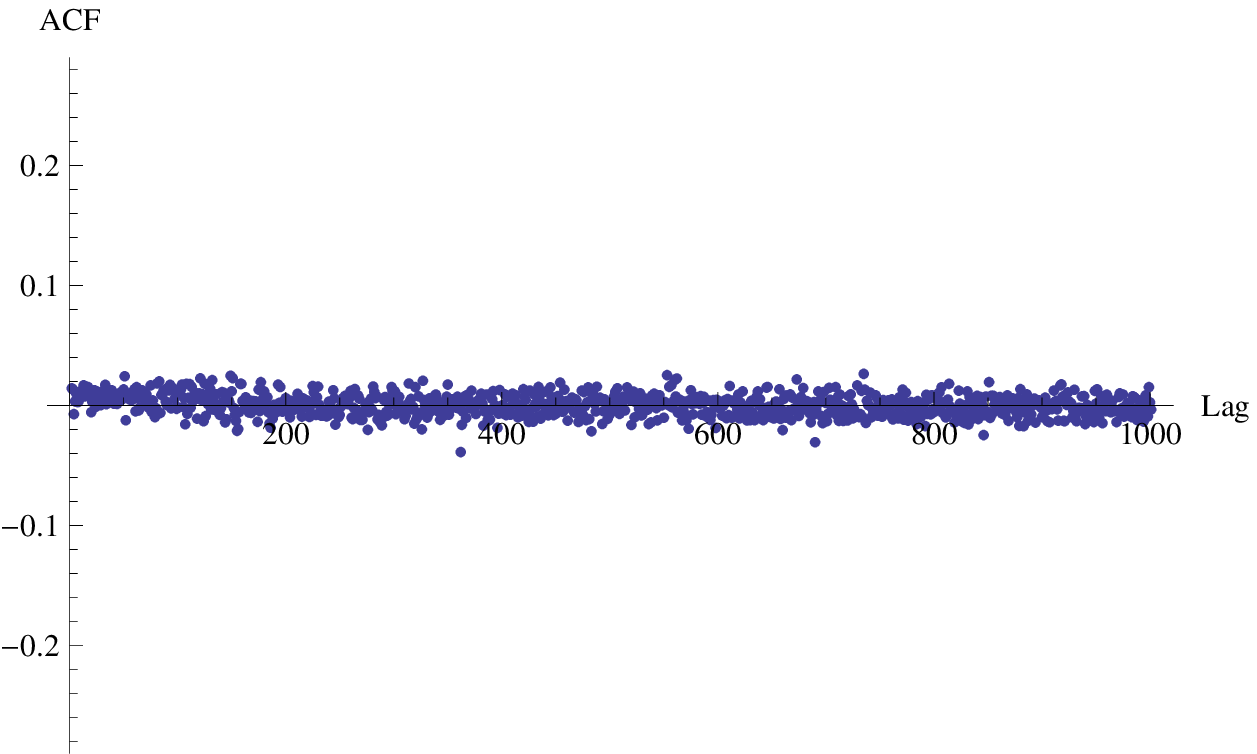}} }
\caption{\label{updraw}ACF for raw and HMM-generated reads (\emph{update-mix} workload)}
\end{figure}

Figure~\ref{updraw} shows the ACFs for reads in the \emph{update-mix} workload, which indicate very little autocorrelation in both cases.

\begin{figure}[!h]
\centerline{
{\setlength{\epsfxsize}{0.5 \hsize}
\epsfbox{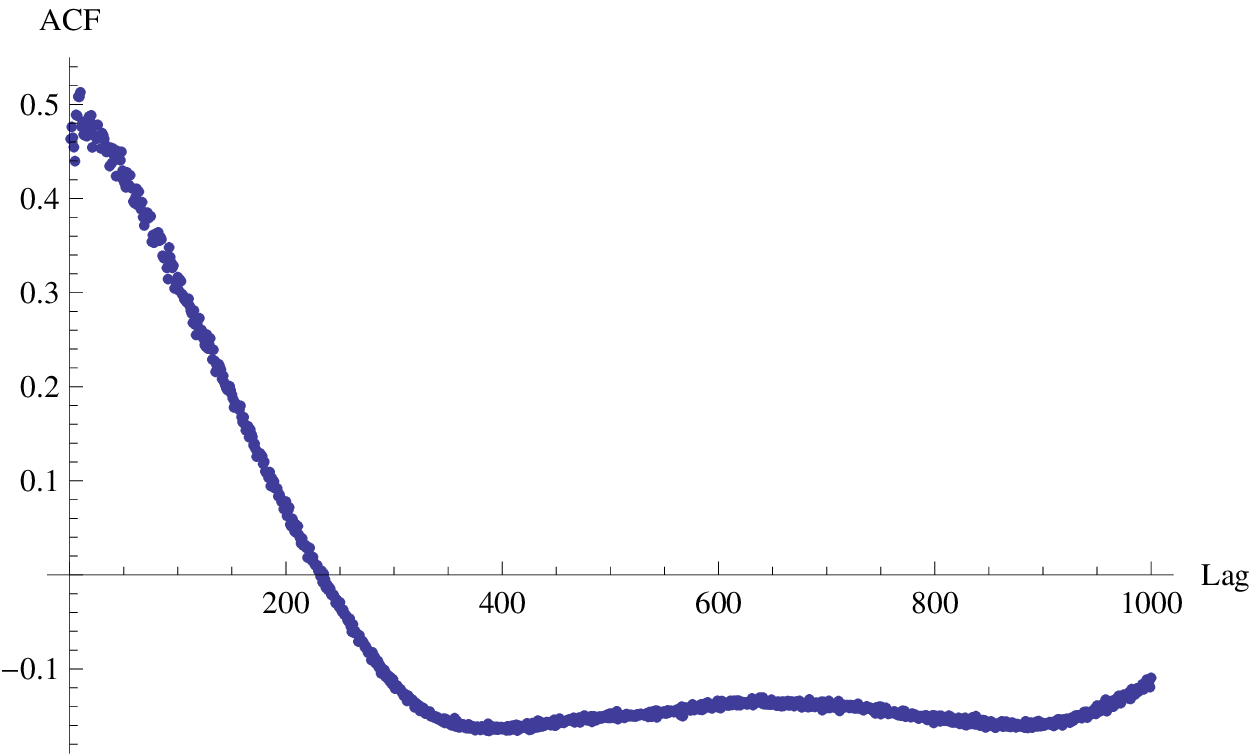}}
\hspace{-0.1in}
{\setlength{\epsfxsize}{0.5 \hsize}
\epsfbox{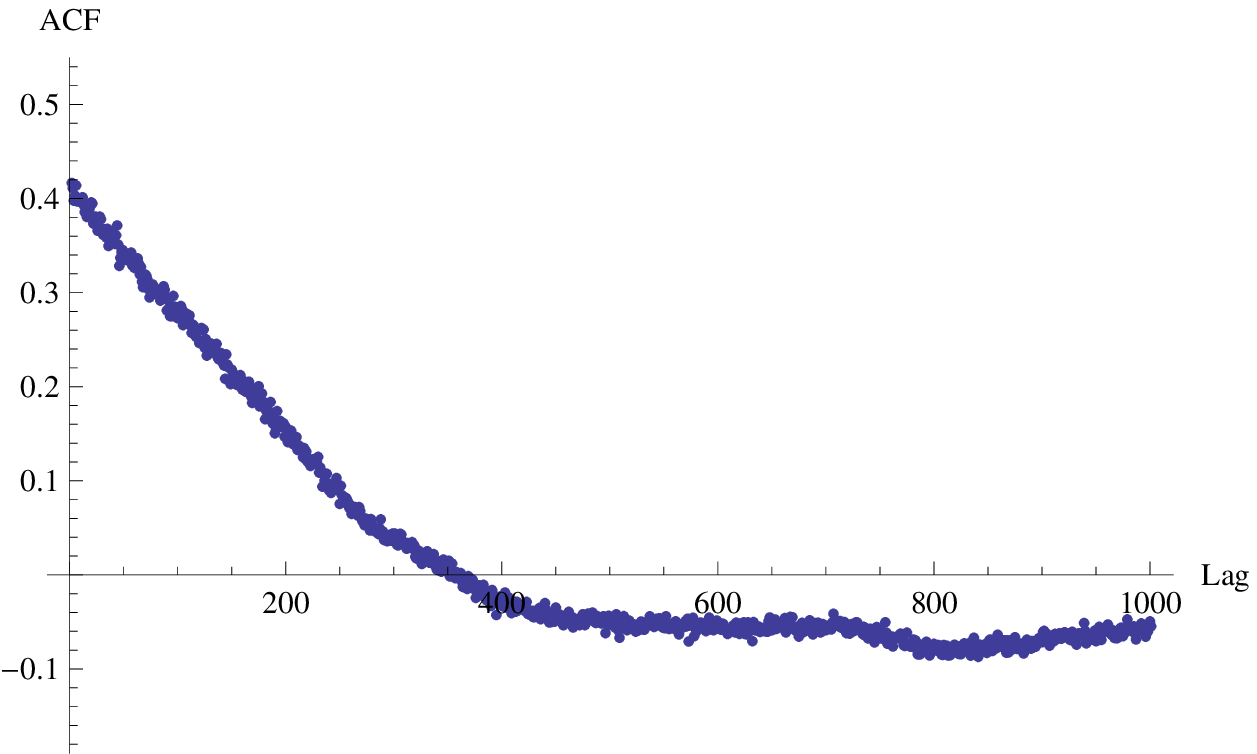}} }
\caption{\label{updHMM}ACF for raw and HMM-generated writes (\emph{update-mix} workload)}
\end{figure}

Figure~\ref{updHMM} shows that writes demonstrate significant autocorrelation -- with qualitatively similar behaviour in both the raw and the HMM-generated traces. This supports the assertion that the HMM-generated output is faithfully reproducing the autocorrelation present in the trace that is used as input.  The magnitude of the HMM's ACF is smaller, probably due to the smoothing arising from the clustering and distributional (Multi-normal) assumptions.

Quantitatively, the HMM-trace matches the raw data trace's ACF fairly well:  starting at about 0.5 at low lags and decreasing, becoming negative at larger lags (between 200 and 400), although with smaller gradient.  At larger lags still (greater than 500) the HMM trace ACF becomes a little larger than that of the raw trace, although remaining negative.  In fact, interpretation at these higher lags is more problematic and the absolute values are, in any event, small in both cases.  Moreover, for different simulated traces using the same HMM, the higher-lag behaviour differs somewhat: in some cases the ACF actually becomes positive again, whilst in other cases the  change from positive to negative occurs at a significantly different lag -- either lower or higher than the lag of about 240 where the raw ACF changes sign.   

\begin{figure}[!h]
\centerline{
{\setlength{\epsfxsize}{0.5 \hsize}
\epsfbox{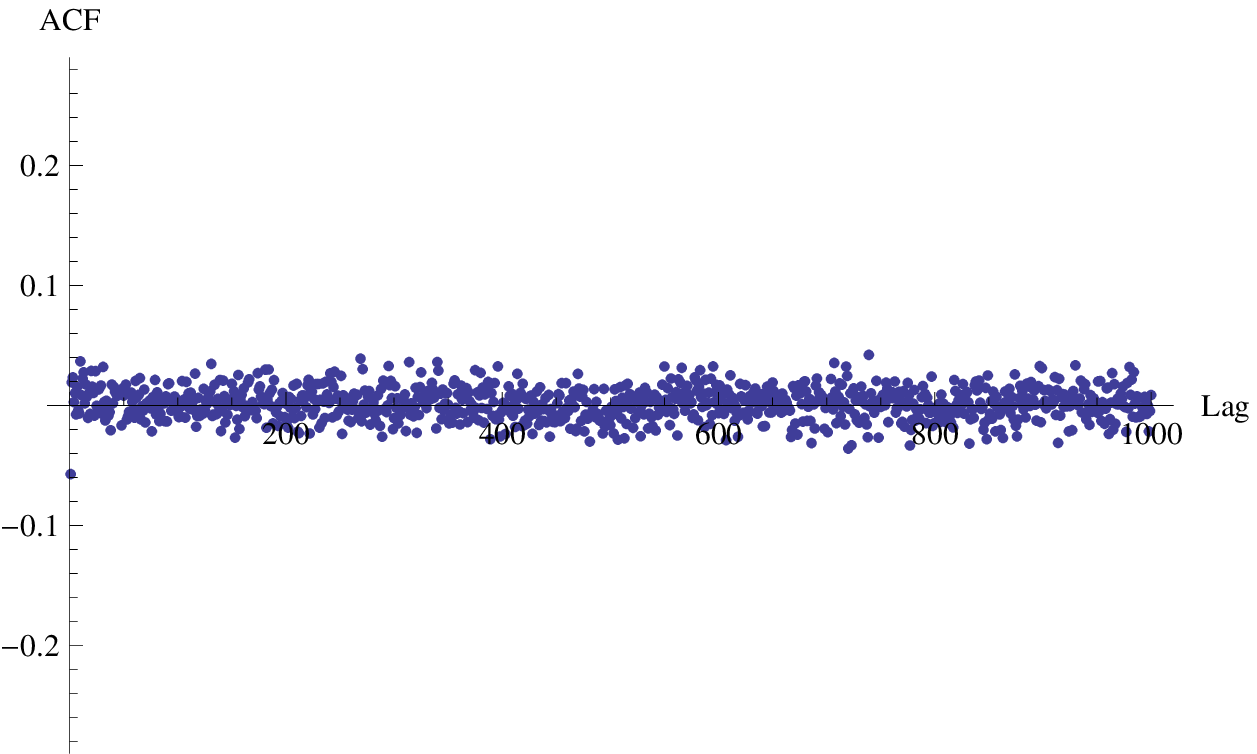}}
\hspace{-0.1in}
{\setlength{\epsfxsize}{0.5 \hsize}
\epsfbox{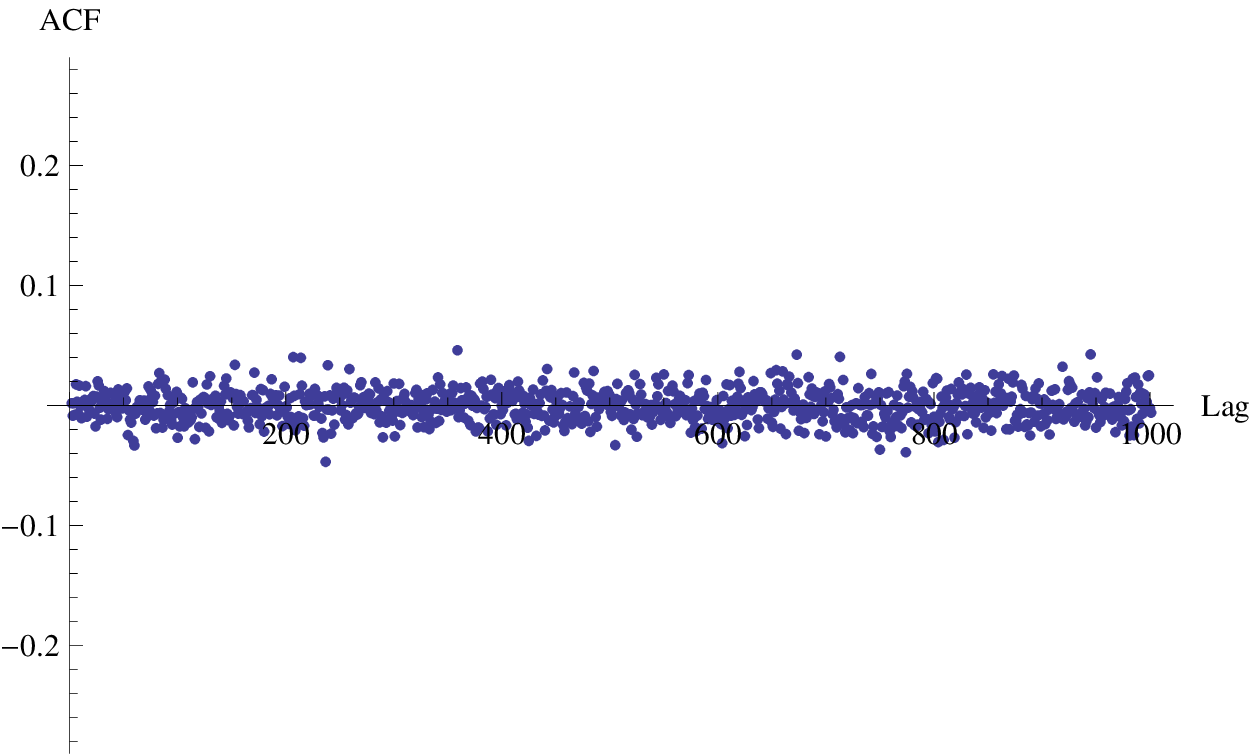}} }
\caption{\label{fig6}ACF for raw and HMM-generated reads (\emph{oltp-mix} workload)}
\end{figure}

Figure~\ref{fig6} shows the ACFs for reads in the \emph{oltp-mix} workload, which comprehensively indicate very little autocorrelation in both cases.

\begin{figure}[!h]
\centerline{
{\setlength{\epsfxsize}{0.5 \hsize}
\epsfbox{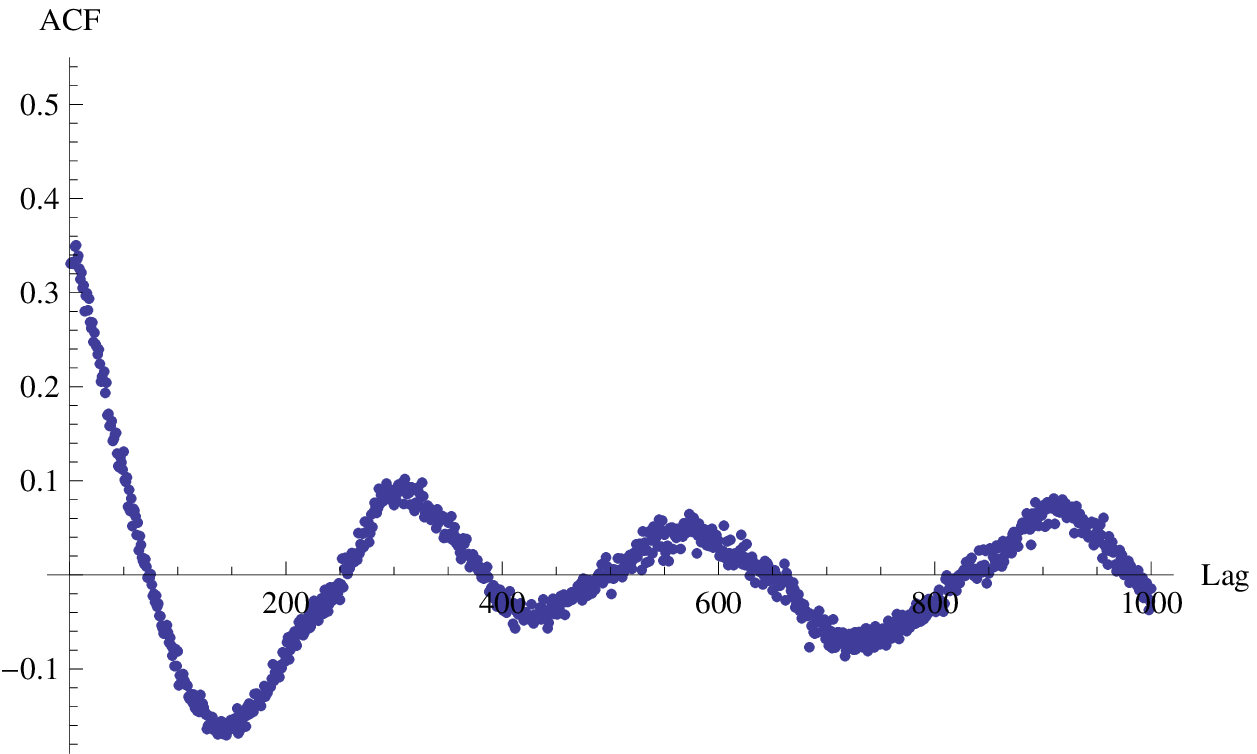}}
\hspace{-0.1in}
{\setlength{\epsfxsize}{0.5 \hsize}
\epsfbox{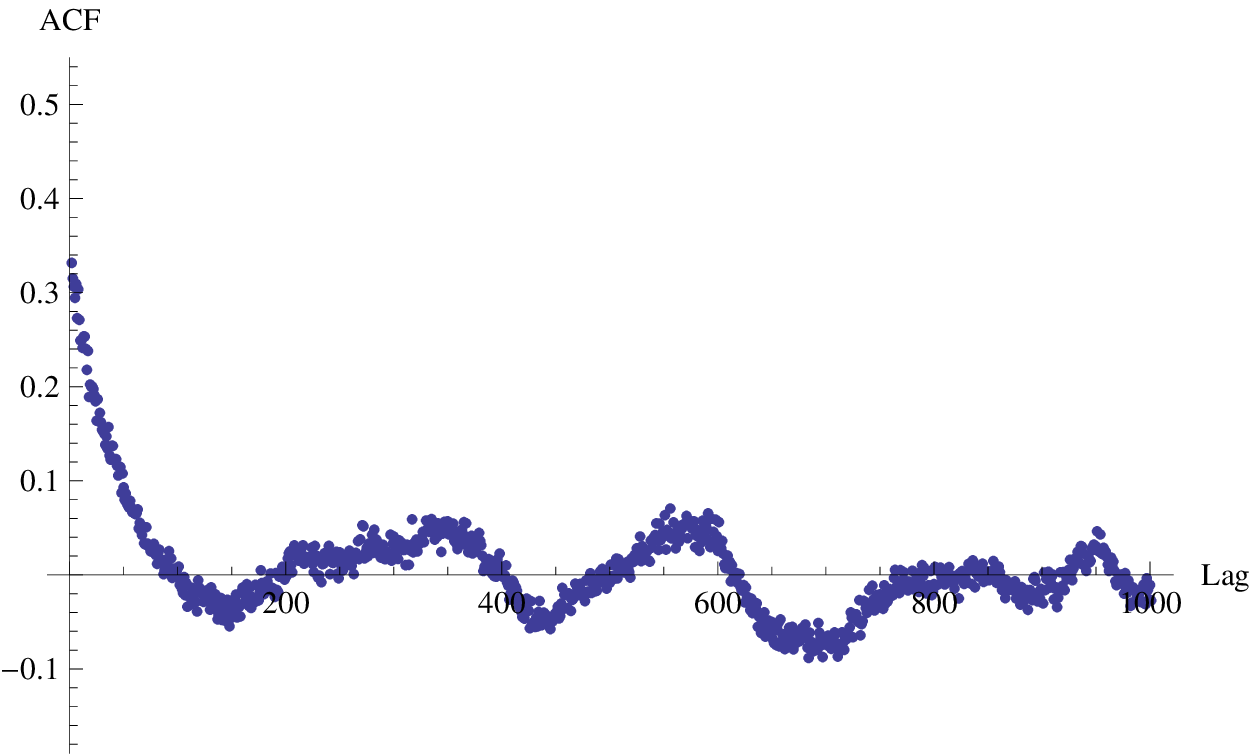}} }
\caption{\label{fig7}ACF for raw and HMM-generated writes (\emph{oltp-mix} workload)}
\end{figure}

Similarly to the \emph{update-mix}, Figure~\ref{fig7} shows that writes again demonstrate significant, and qualitatively similar, autocorrelation in both the raw and the HMM-generated traces, although of a rather different nature. This supports the assertion that the HMM-generated output is again faithfully reproducing the autocorrelation present in the trace that is used as input.

It is noteworthy that the HMM trace matches the oscillating pattern of the write-ACF even at large lags, although the magnitude is a little smaller, probably because of the aforementioned smoothing. 

Finally, it turned out that there was little (but significantly non-zero) correlation between the numbers of reads and the numbers of writes in each of the workload mixes -- a positive correlation in the case of the  \emph{update-mix} and negative for the  \emph{oltp-mix}.  The agreement between the raw and HMM traces was good:  respectively $0.025$ and $0.024$ for the  \emph{update-mix}; $-0.081$ and $-0.072$ for the \emph{oltp-mix}.  

\subsubsection{Queueing times}
We already have a Flash hardware simulator~\cite{peva10} and we ran it
using, first, the same raw binned traces (\emph{update-mix} and \emph{oltp-mix}) as above, 
then corresponding HMM-generated traces.   The
mean queueing times for each of the three operation types (read/write/erase) 
were then estimated from the simulation runs (using batch-means with five runs per batch) for three 
different priority schemes: (\emph{i}) strict first come, first served (FCFS)
queueing (``no priority''); (\emph{ii}) FCFS with non-preemptive read-priority;
and (\emph{iii}) FCFS with preemptive read-priority.  

 \begin{table} 
 {\footnotesize 
\begin{tabular}{|c|c||c|c|c|c|c|c|} \hline
 \multicolumn{2}{|c||}{Scenario} & \multicolumn{3}{c|}{Raw~~}  & \multicolumn{3}{c|}{HMM~~}  \\ \hline 
 Scheme & Workload & Read & Write & Erase & Read & Write & Erase \\
\hline \hline
No & \emph{update-mix} &  5.29 & 6.46 & 6.46 & 4.23$\pm$ 1.03 & 5.40 $\pm$ 1.56 & 5.40$\pm$ 1.56 \\
priority & \emph{oltp-mix} & 17.25 & 16.99 & 17.00 & 19.63$\pm$ 2.61 & 19.38$\pm$ 3.17 & 19.36 $\pm$ 3.18\\
\hline 
Non-pre- & \emph{update-mix} & 0.76  & 7.12 & 7.11 & 0.38$\pm$ 0.03 & 6.03$\pm$ 1.64 & 6.02$\pm$ 1.64 \\
emptive & \emph{oltp-mix} & 0.0006 & 30.41 & 30.42 & 0.006$\pm$ 0.001 & 34.66$\pm$ 5.90 & 34.64$\pm$ 5.92\\
\hline 
Pre- & \emph{update-mix} & 0.76  & 7.12 & 7.11 & 0.10$\pm$ 0.01 & 6.03$\pm$1.64 & 6.03$\pm$ 1.64  \\
emptive & \emph{oltp-mix} & 0.0003 & 30.41 & 30.42 & 0.005$\pm$ 0.002 & 34.66$\pm$ 5.90 & 34.65$\pm$ 5.92\\
\hline
\end{tabular} }
\centering{
\caption{ \label{table2} Mean queueing times (ms) for each workload type, for raw and HMM traces, with 95\% confidence intervals}
}
\end{table}

These cases yielded the comparisons shown in Table~\ref{table2}.  The mean queueing times for write and erase memory accesses predicted using the HMM traces agree well with the corresponding results produced using the raw traces, lying well inside the 95\% confidence bands\footnote{In fact the erase values are rather redundant since the raw traces did not include erases, which are added post-WAFL by a FTL.  In our simulator, we scheduled one erase per 64 writes, on average, to free space at the right rate (one erase-block = 64 write-pages); for more details see~\cite{peva10}.  Thus, if the write results agree, the erase results must also.}.   In fact the HMM traces lead to underestimates for the \emph{update-mix} and overestimates for the  \emph{oltp-mix}.  For reads, the agreement appears poor, except in the no-priority case, where it is satisfactory with a queueing time of about one bin-width (5ms) on average.   For non-preemptive priority, there is a 50\% error in the case of the \emph{update-mix} and an order of magnitude error for the  \emph{oltp-mix}.  The error is worse with preemptive priority.  However, it must be remembered that here the mean queueing times are much less than a bin-width and so the model's granularity of about 5ms is too coarse to give accurate predictions.  In fact the raw trace data give a mean read queueing time of about 15\% of a bin-width for the  \emph{update-mix} and less than 0.1\% of a bin-width for the  \emph{oltp-mix}.   The HMM again underestimates in the former case and overestimates in the latter.  It is likely that the error arises from the assumed bivariate Normal distribution of read-write data-points within the clusters; in practice, the points are likely to be skewed, e.g. with a larger number of very small interarrival times in the case of the  \emph{oltp-mix}, causing greater contention.   

This error is actually unimportant since it is clear that the queueing times are negligible for reads in both priority schemes and can reasonably be taken as zero.  However, we did investigate further by aggregating the time-stamped data into 1ms bins.  This had the effect of producing a significant number of empty bins, with no reads or writes; about 20\% of all observations, in fact.  The HMM inferred that the empty bins constituted a new state that turned out to dominate the model, which no longer represented the other hidden states so well.  The mean read queueing times did improve considerably under the two priority schemes, but the bin size was still too small to gain significantly more accuracy; a continuous time version of the HMM would be needed.   The numerical results for the 1ms bin size are not reproduced here since they add little and have no practical value.

\section{Real user workload} \label{user-workload}
The preceding analysis concerned a synthetic user workload that had been transformed to enhance its efficient access to Flash devices. However, it is equally important to be able to model similarly the real user workloads that are submitted externally to a networked storage system, be it Flash-based, disk-based, a combination of both or other technologies. The HMM-based approach described in this paper is intended for all such workload types and, indeed, other kinds of time series, arising in communication, financial and logistic systems, for example. We therefore tested the methodology on a real user workload comprising CIFS (Common Internet File System) transactions monitored on a production NetApp system.  These time-stamped traces are anonymized aggregates of the actual storage accesses requested by a number of users over periods of up to 100 minutes and were collected using the \emph{tcpdump} command line packet analyzer.  The intensity of this workload is very much greater than that analyzed in the previous section since it represents many users' accesses to a whole storage system, whereas the former workload relates to a single Flash device.  Out of a suite of about 1000 traces, we selected the largest one, corresponding to highest intensity -- the monitored time period being close to 100 minutes for all traces.

Following precisely the procedure described in section~\ref{model} and illustrated in Figure~\ref{fig4}, we first assigned the entries of the chosen trace to bins of width one second and applied the same, Mathematica$^{\tiny\tt TM}$ clustering algorithm as in section~\ref{HMMwlmodel}.  There are many ``empty'' bins -- bins with no reads and no writes -- and so, in addition to the $15$ clusters identified by the clustering algorithm, we also defined a singleton cluster consisting of the point $(0,0)$.   We then ran our Baum-Welch algorithm, with four hidden states, to obtain estimates for the parameters $Q,G,\nu$ of the proposed HMM.  The four states turned out to represent empty bins, and, similarly to the preceding analyses, predominantly small reads and writes, with and without significant medium sized reads and writes, and a state with a few very large writes.  From this we generated traces of cluster identifiers, by simulation of the HMM, and hence numbers of read block-counts and write block-counts in a sequence of 1s bins of length equal to that of the raw trace; we again used truncated multi-normal distributions with parameters given by the means, variances and covariance of the reads and writes within the clusters.  This resulted in the excellent agreement shown in Table~\ref{tableCIFS} for the means and standard deviations of the whole traces.  All raw values lie well within the 95\% confidence bands derived from batch-means simulation with 10 batches.  Notice, however, the very wide bands on the standard deviations.  These are due to the high values of the standard deviations which come from the presence of clusters with a very wide range of centroid values: the smallest for reads (respectively writes) is $0 (0)$ (because of the singleton cluster) and the largest is  $5.278 \times 10^6 (7.82 \times 10^6)$.  The correlations between the read and write block-counts are $0.24$ for the raw trace and $0.23$ for the aggregate of the HMM traces used in the batch-means procedure.

 \begin{table} 
 {\footnotesize 
\begin{tabular}{|c||c|c|c|c|c|c|c|c|} \hline
 & \multicolumn{4}{c|}{Reads/bin} & \multicolumn{4}{c|}{Writes/bin} \\ \hline 
 Workload & \multicolumn{2}{c|}{Raw~~~}  & \multicolumn{2}{c|}{HMM~~~} & \multicolumn{2}{c|}{Raw~~~} & \multicolumn{2}{c|}{HMM~~~}  \\ \hline
type & Mean & Std Dev  & Mean & Std Dev & Mean & Std Dev  & Mean & Std dev \\
\hline \hline 
\emph{1s bins} & 86,805 & 220,471 & $\begin{array}{c} 92,187 \\ 81,744 \end{array}$ & $\begin{array}{c} 242,467 \\ 196,031 \end{array}$  & 11,653 & 201,281 & $\begin{array}{c} 14,347 \\ 11,653 \end{array}$ & $\begin{array}{c} 247,294 \\ 173,872 \end{array}$  \\
\hline 
\emph{100ms bins}  & 12,346 & 45,547 & $\begin{array}{c} 12,548 \\ 11,932 \end{array}$ & $\begin{array}{c} 45,975 \\ 44,526 \end{array}$  & 955 & 11,193 & $\begin{array}{c} 1,026 \\ 923 \end{array}$ & $\begin{array}{c} 12,302 \\ 10,378 \end{array}$ \\
\hline
\emph{Reduced bins}  & 30,856 & 67,927 & $\begin{array}{c}33,104 \\ 28,611 \end{array}$ & $\begin{array}{c} 70,484 \\ 66,046 \end{array}$  & 2,386 & 17,600 & $\begin{array}{c} 3,521 \\ 1,563 \end{array}$ & $\begin{array}{c} 21,985 \\ 15,740 \end{array}$ \\
\hline
\end{tabular} }
\centering{
\caption{ \label{tableCIFS} Bin means and standard deviations for CIFS traces, with 1s bins and 95\% confidence intervals for the HMM-generated traces.}
}
\end{table}

The autocorrelation functions also show good agreement, in fact rather better than in the comparisons of section~\ref{tests}.  However, there appears to be negligible autocorrelation for either reads or writes.  As a result, we undertook a finer grained analysis.

\begin{figure}[!h]
\centerline{
{\setlength{\epsfxsize}{0.5 \hsize}
\epsfbox{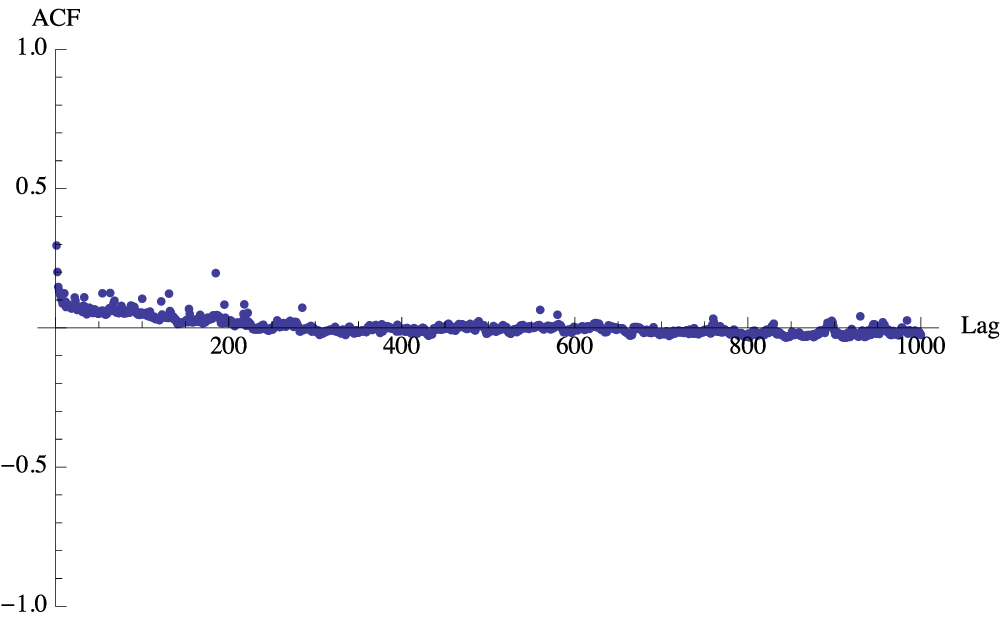}}
\hspace{-0.1in}
{\setlength{\epsfxsize}{0.5 \hsize}
\epsfbox{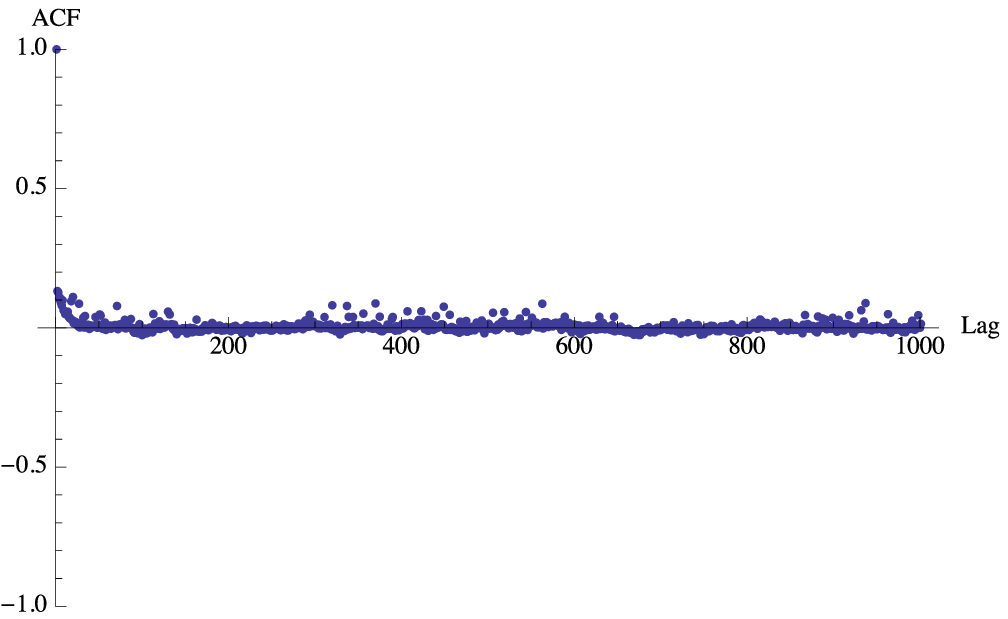}} }
\caption{\label{CIFS1s_reads}ACF for raw and HMM-generated reads (CIFS workload with 1s bins)}
\end{figure}

\begin{figure}[!h]
\centerline{
{\setlength{\epsfxsize}{0.5 \hsize}
\epsfbox{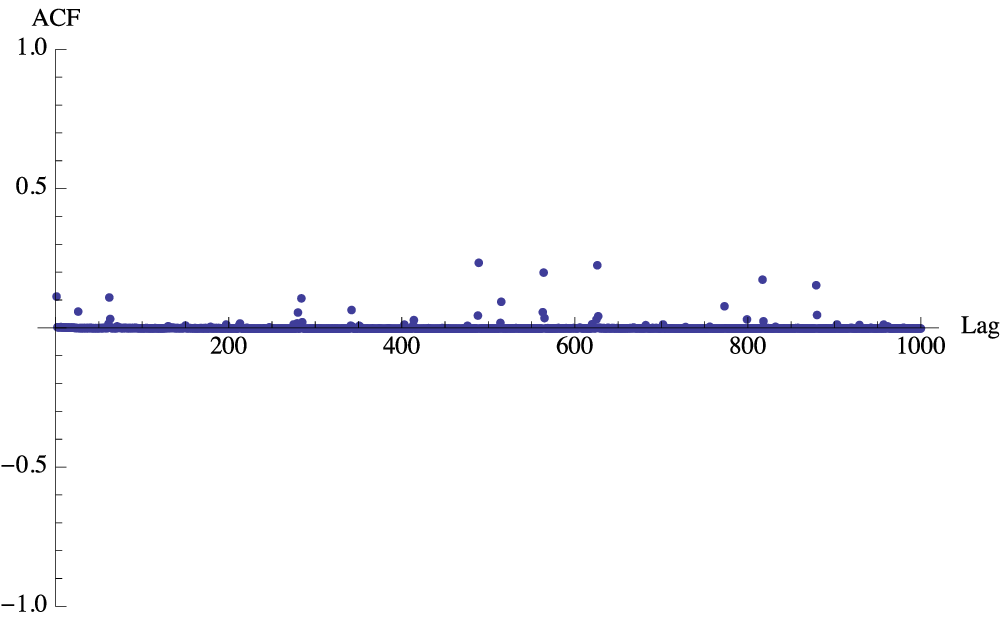}}
\hspace{-0.1in}
{\setlength{\epsfxsize}{0.5 \hsize}
\epsfbox{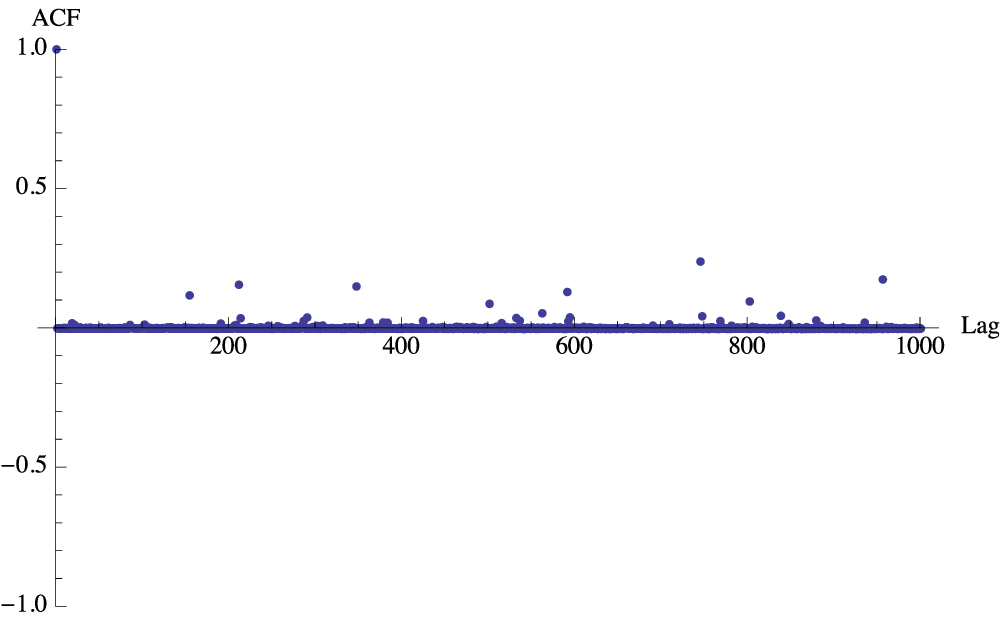}} }
\caption{\label{CIFS1s_writes}ACF for raw and HMM-generated writes (CIFS workload with 1s bins)}
\end{figure}

\subsection{Finer grain analysis} \label{cifs100ms}
We increased the detail of our analysis by a factor of ten by dividing the time-stamped traces into $100$ms bins.  Smaller bins still, e.g. 10ms, gave very many empty bins and few significant sequences of non-empty bins; this suggested there would be no useful hidden state dynamics for discrete observations, which would be dominated by the empty category.  Table~\ref{tableCIFS} again reveals excellent agreement on the basic statistics for the whole traces, and the correlations between reads and writes were extremely close at $0.140$ for the raw trace and $0.144$ for the HMM trace.  Turning to the autocorrelations, we found some surprising results.

\begin{figure}[!h]
\centerline{
{\setlength{\epsfxsize}{0.5 \hsize}
\epsfbox{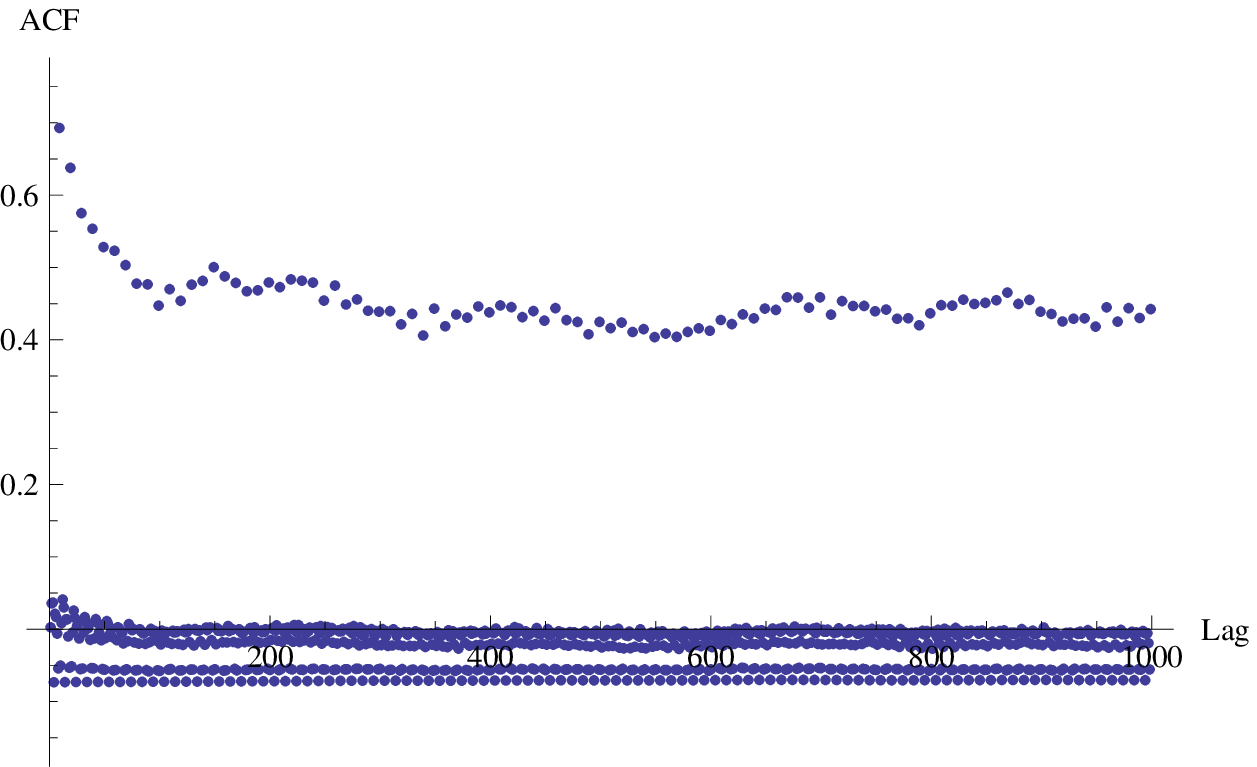}}
\hspace{-0.1in}
{\setlength{\epsfxsize}{0.5 \hsize}
\epsfbox{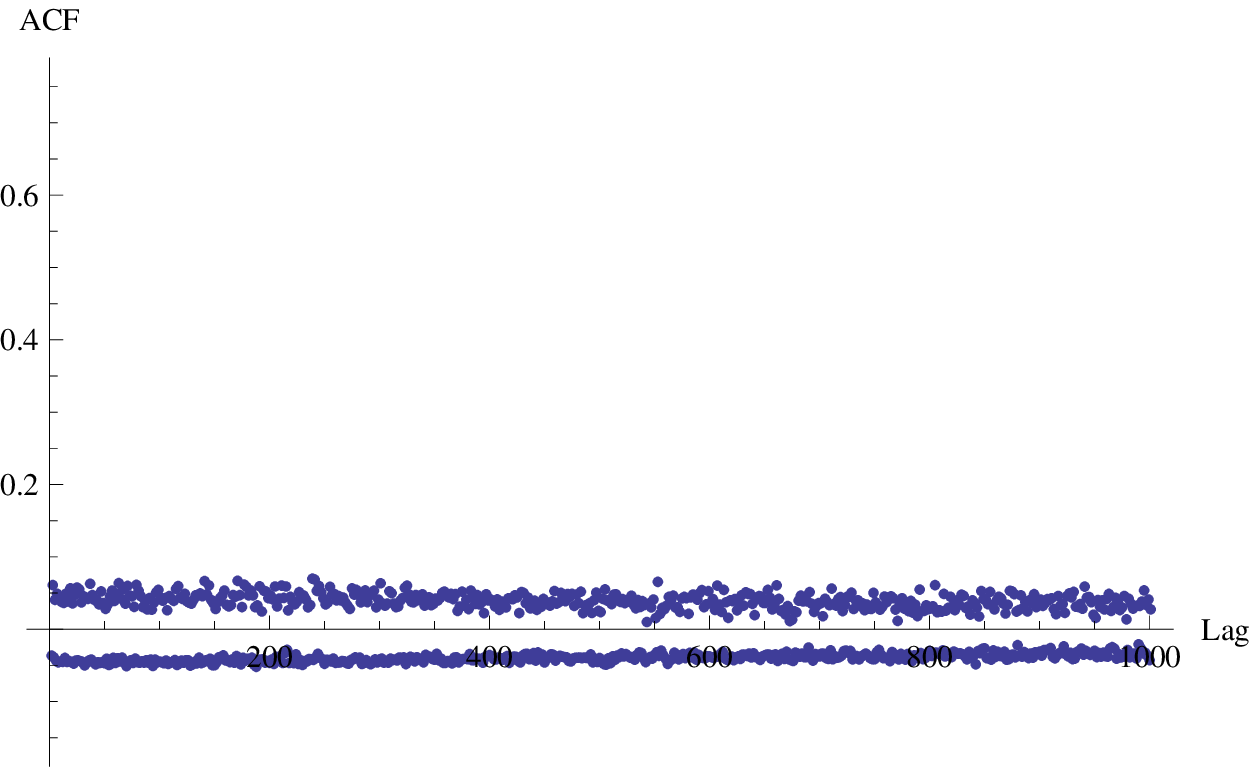}} }
\caption{\label{CIFS100ms_reads}ACF for raw and HMM-generated reads (CIFS workload with 100ms bins)}
\end{figure}

\begin{figure}[!h]
\centerline{
{\setlength{\epsfxsize}{0.5 \hsize}
\epsfbox{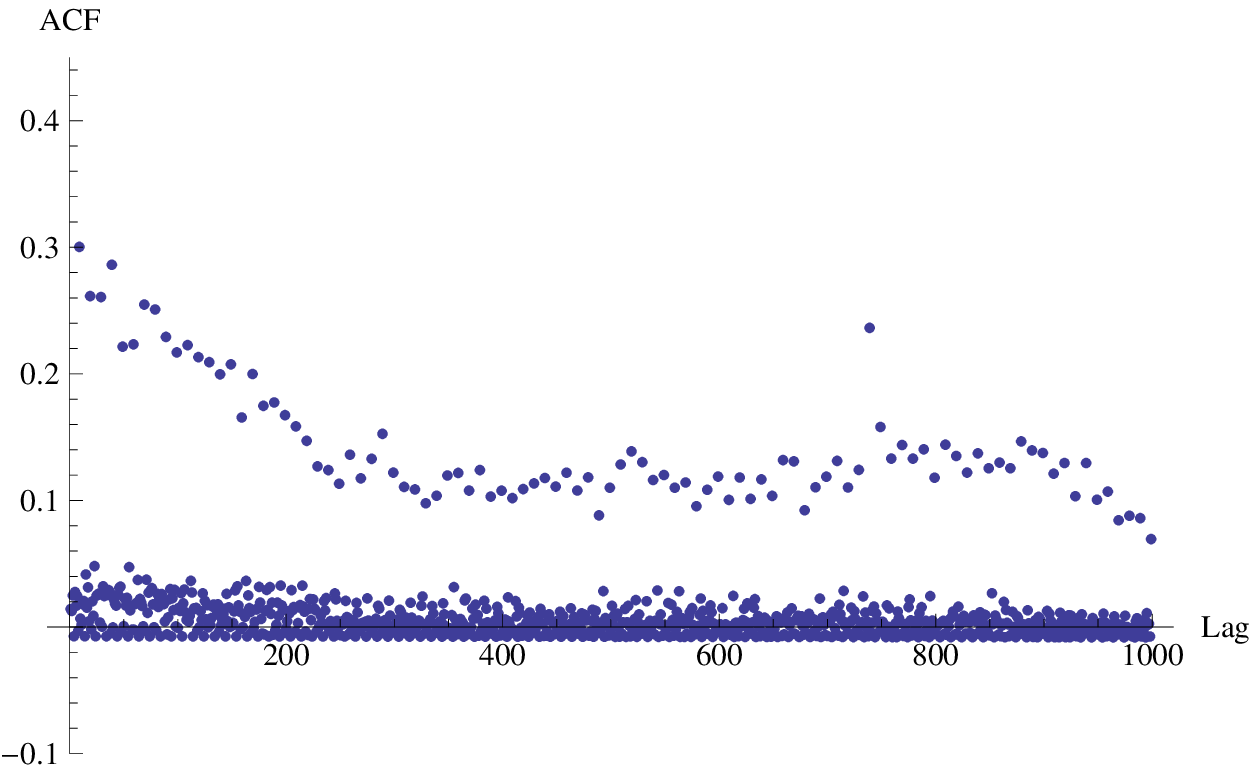}}
\hspace{-0.1in}
{\setlength{\epsfxsize}{0.5 \hsize}
\epsfbox{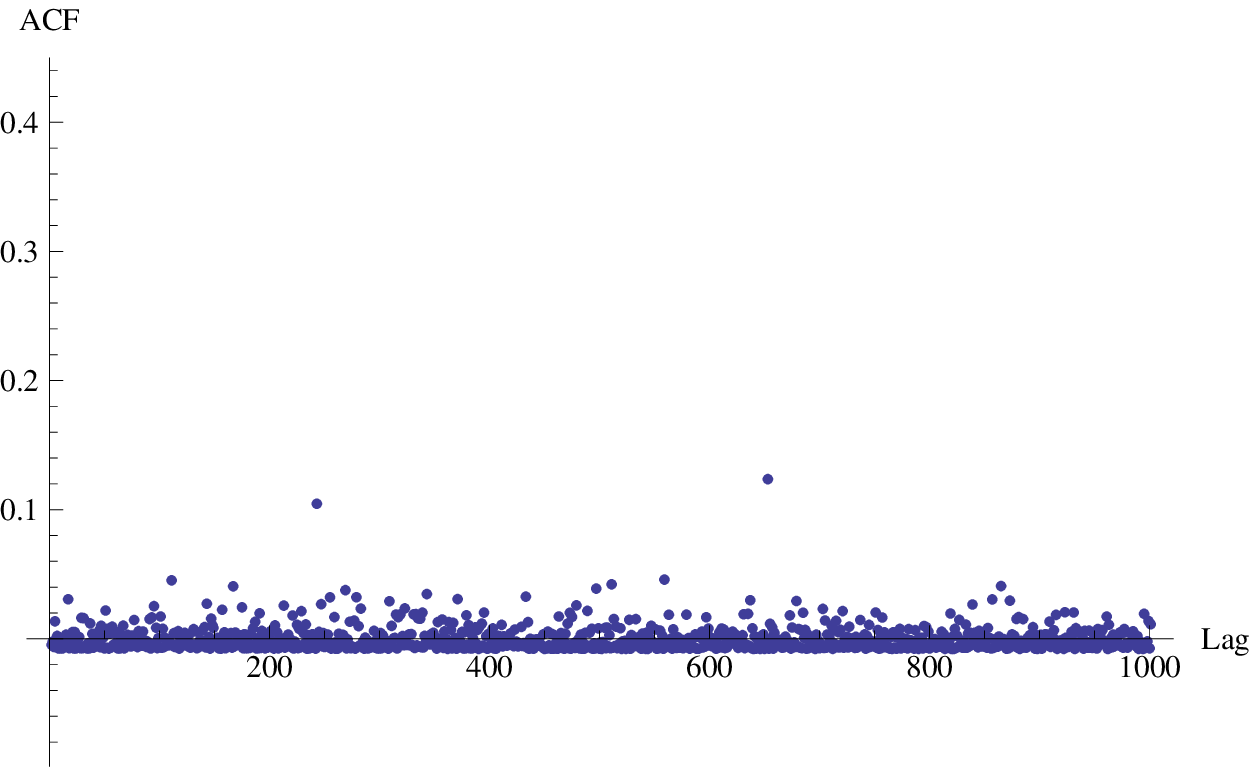}} }
\caption{\label{CIFS100ms_writes}ACF for raw and HMM-generated writes (CIFS workload with 100ms bins)}
\end{figure}

For the raw data, over a range of lags up to 1000, the \emph{appearance} is of two plots (Figures~\ref{CIFS100ms_reads} and~\ref{CIFS100ms_writes}, left graphs), one showing negligible correlation and the other showing significant correlation decreasing from lag 1, where it is about 0.7 for reads and 0.3 for writes.  For the reads, even the low correlation part appeared in two bands, which was matched well in the HMM autocorrelation function for reads (see Figure~\ref{CIFS100ms_reads}.   Essentially, the HMM agreed well with the lower plots in each case but did not account for the higher plot.

\begin{figure}[!h]
\centerline{
{\setlength{\epsfxsize}{0.5 \hsize}
\epsfbox{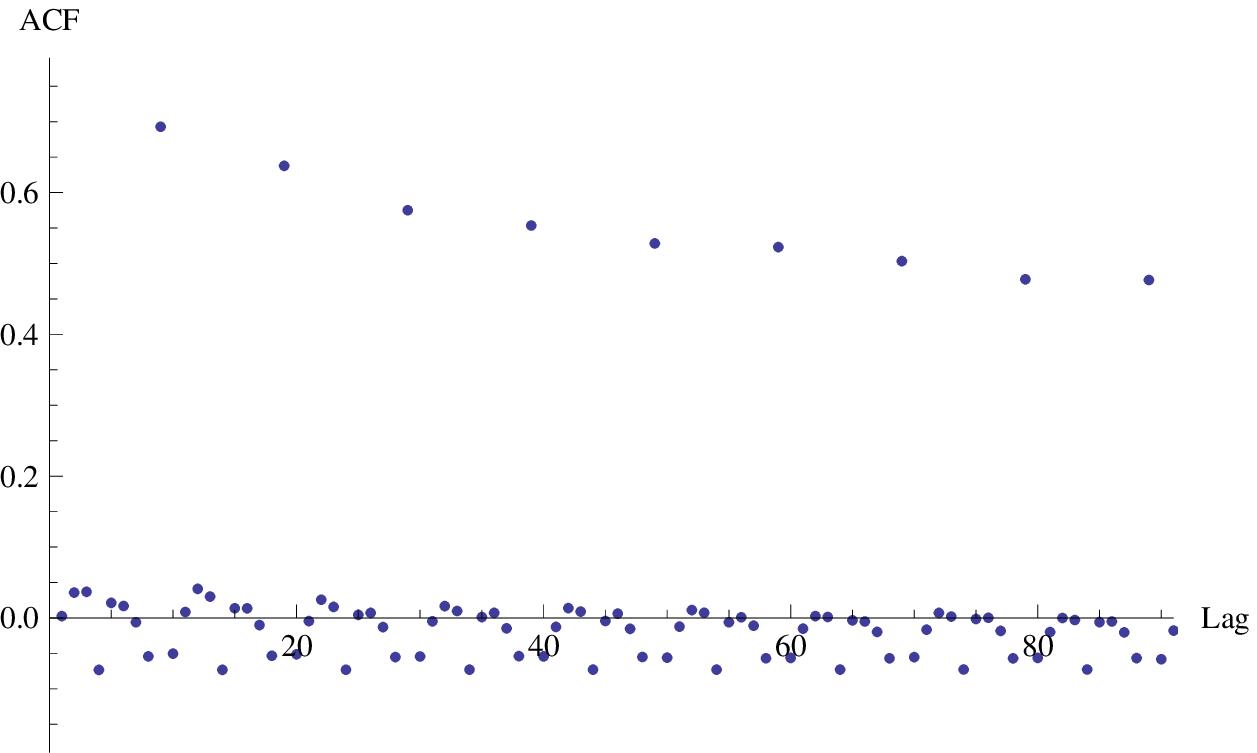}}
\hspace{-0.1in}
{\setlength{\epsfxsize}{0.5 \hsize}
\epsfbox{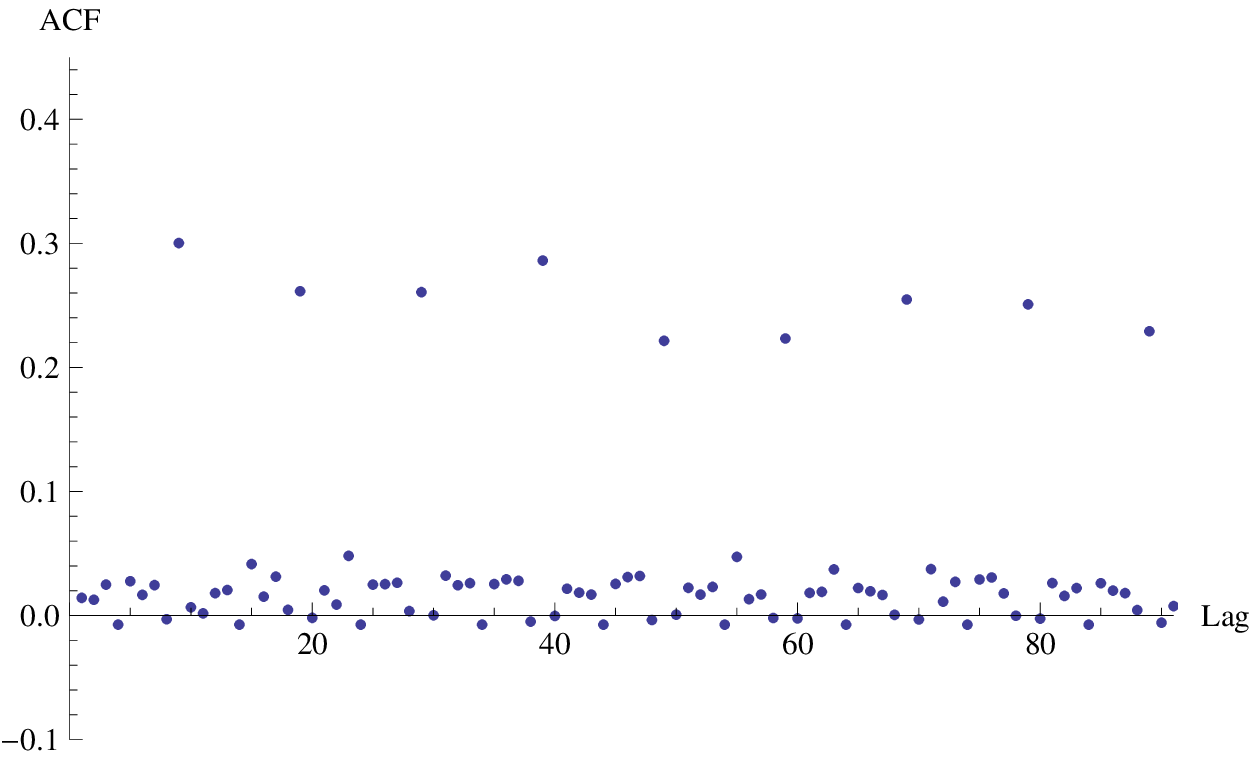}} }
\caption{\label{CIFS100ms_0-91}  Exploded views of the periodic, raw ACFs for 100ms binned traces at lags 1 -- 91. Left: reads, Right: writes. }
\end{figure}

The cause for this behaviour in the raw workload is an unexpected periodicity.  Figure~\ref{CIFS100ms_0-91} shows the raw autocorrelations for lags up to 91 and reveals that the period is 10 lags, with the upper plot occurring only at lags $10, 20, 30, \ldots$.  With a HMM based on just four hidden states, it is not surprising no periodicity was found in the HMM traces\footnote{We also fitted a 12-state HMM to the observation trace in order to verify that a HMM could actually account for this behaviour.  In practice this would be an unparsimonious, somewhat pointless exercise since there would be almost as many hidden states as observation values (12 versus 16) and it would be extremely difficult to interpret each individual state physically.  Nevertheless, we show the results of this experiment in Appendix~\ref{B}.}.

\subsection{Periodic time series}
There is no apparent reason for such periodic behaviour, but we noted that the period in real time is exactly one second and then further investigated the individual bin contents in the 100ms binned trace.  We discovered that the second half of every one second bin was empty, as indeed was the fourth 100ms segment.  Thus, out of every second, the only possible non-empty 100ms bins were indexed $1,2,3,5$.  This may be due to the time-stamp recording in \emph{tcpdump} or, in the case of the second halves of the 1s bins, IO scheduling to avoid saturation of the storage devices.  Another possibility is, of course, an error in the recording of the time-stamps.  We did not investigate the cause further but did continue our analysis by deleting all the inevitably empty 100ms bins -- those indexed $4,6,7,8,9,10$ in each second -- reducing the length of the trace to 40\% of the original.   We increased the number of hidden states in the HMM to six, to better facilitate finding periodic behaviour with period 4.  The aggregate results again showed good agreement -- see Table~\ref{tableCIFS} -- and the correlations between reads and writes were $0.112$ and $0.109$ for the raw and HMM traces respectively.

\begin{figure}[!h]
\centerline{
{\setlength{\epsfxsize}{0.5 \hsize}
\epsfbox{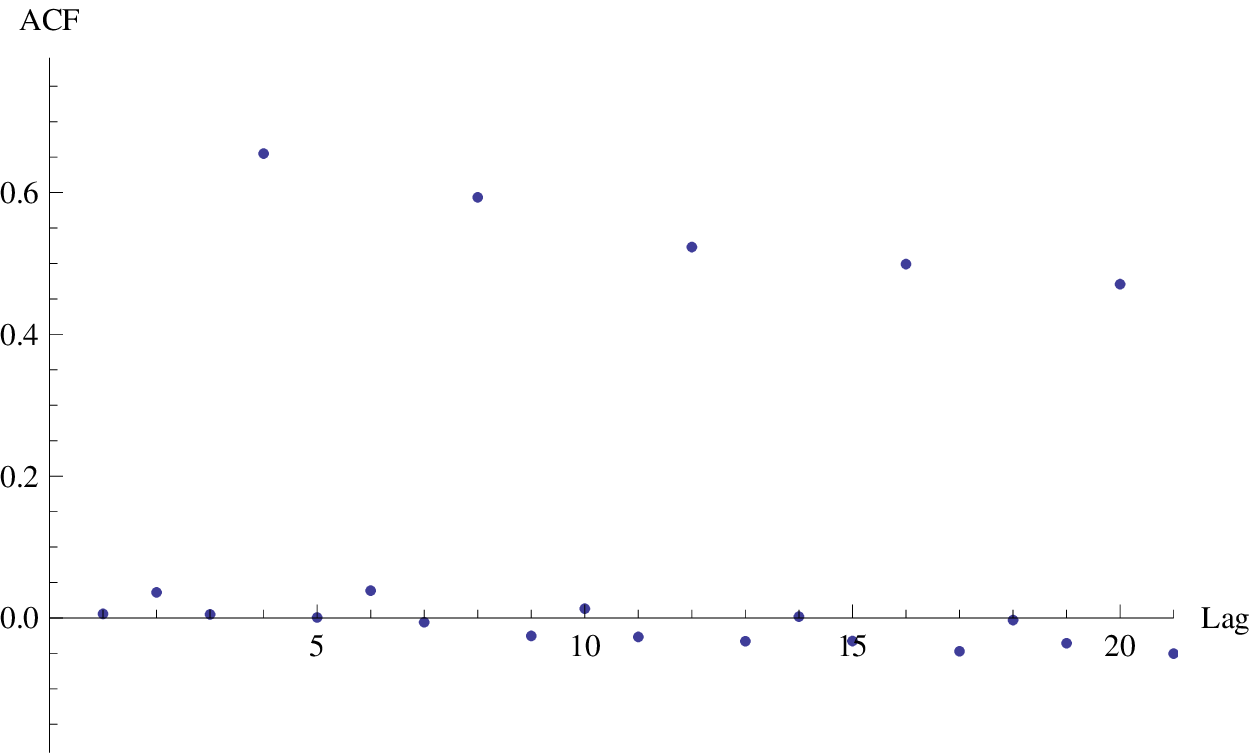}}
\hspace{-0.1in}
{\setlength{\epsfxsize}{0.5 \hsize}
\epsfbox{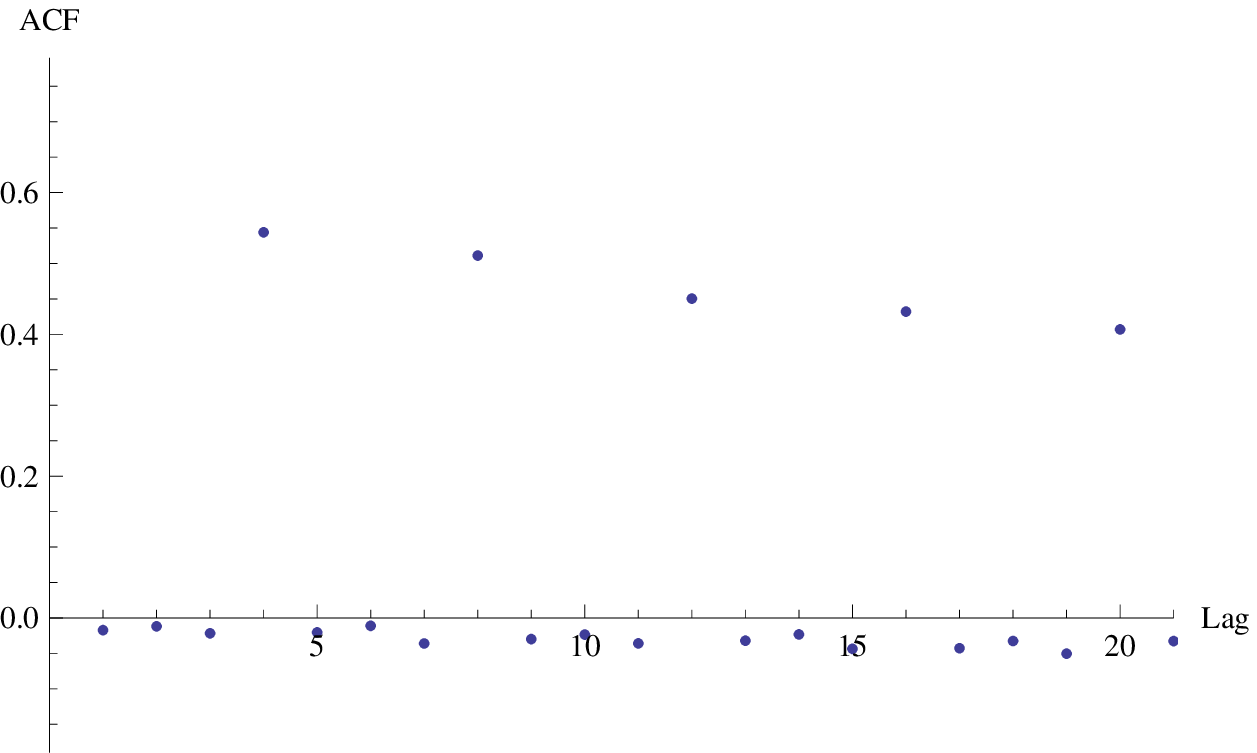}} }
\caption{\label{CIFSred_reads_1-21}ACF for raw and HMM-generated reads at lags 1 -- 21 (thinned 100ms bins)}
\end{figure}

\begin{figure}[!h]
\centerline{
{\setlength{\epsfxsize}{0.5 \hsize}
\epsfbox{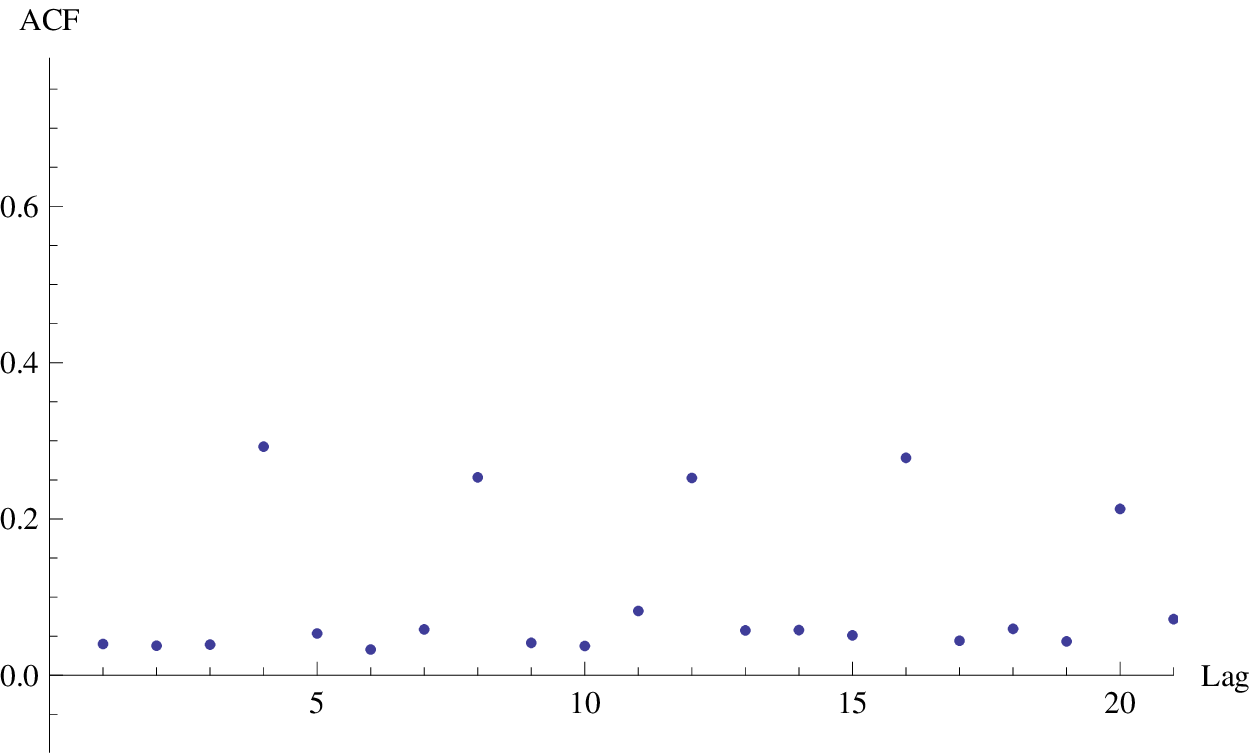}}
\hspace{-0.1in}
{\setlength{\epsfxsize}{0.5 \hsize}
\epsfbox{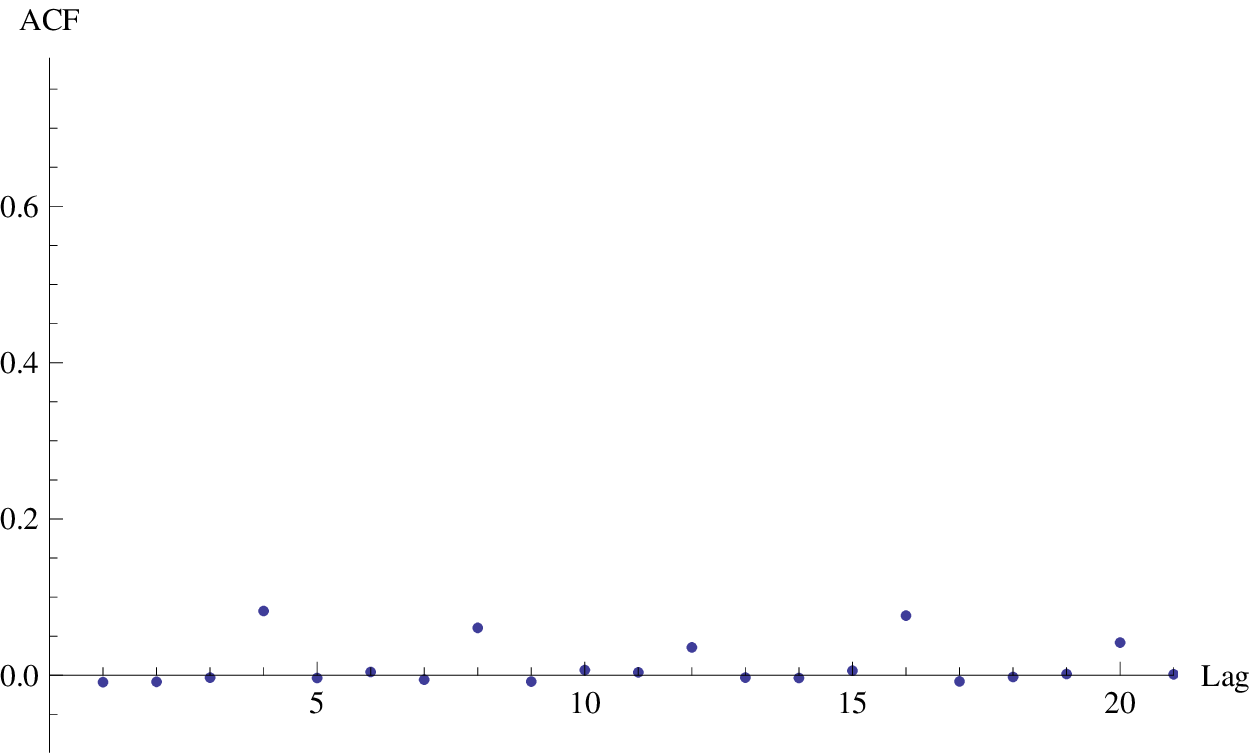}} }
\caption{\label{CIFSred_writes_1-21}ACF for raw and HMM-generated writes at lags 1 -- 21 (thinned 100ms bins)}
\end{figure}

The raw ACFs were still periodic and qualitatively similar to those for the 100ms binned trace, but with period 4; see Figures~\ref{CIFSred_reads_1-21} and~\ref{CIFSred_writes_1-21}, which show exploded views of the ACF up to lag 21.  Pleasingly, the HMM showed similar periodicity -- at a lower amplitude (especially for writes) but certainly with the correct period of 4.  This attests to the great flexibility in our modelling approach, since we did not choose HMMs with a view to accounting for periodic behaviour, which we were not expecting, and yet we were able to predict it faithfully.  We can still only speculate as to the precise cause -- probably arising in the recording of real time -- but this, we feel, is unimportant here; the workload modelling approach itself was proved successful.



\section{Flash workload-performance fluid model}
\label{FlashModel}
Whilst the Flash device itself is relatively simple, the design
choices for higher levels of the storage stack create a wide range of
possible performance capabilities.  Although the randomness or 
sequentiality of read accesses is not an issue for Flash (unlike hard disk storage), the
extent to which the randomness of write accesses affects programs' performance is crucial.  
Higher level file systems such as WAFL (Write Anywhere File Layout) and ZFS (Zettabyte File System) 
may significantly reduce the impact of random writes.  However, even then, 
there is increased erase traffic and likely to be some ``write amplification'', where  
extra writes (and also reads) are needed for the copying done to release whole 
memory blocks, especially to accommodate (logical) in-place updates.  

The flow of traffic from the user to the storage stack is given by a workload model of the 
IO streams entering the FTL layer.  In the case of a fluid model, the modulating 
Markov process controls the fluid input rate 
in each of its phases.  These phases are considered as the hidden states of a HMM workload model, 
from which the modulating chain's generators are obtained.  Our ability to collect traces at a low level 
in the storage stack but inability to measure at the Flash device itself requires some approximations.  
Whenever possible, the FTL design parameters (from vendors) should be used to modify the 
workload-specific rates to comply with the hardware input rates (i.e. the FTL output).  
The Flash model also 
handles the additional erases that are needed to maintain the sustained page write rates.

The HMM-generated traces, representing particular monitored workloads, provide input to an existing, fluid-based Flash performance model~\cite{peva10}.  This model uses a Markov modulated fluid input process, for which we determine the one-step transition probabilities directly from the Baum-Welch estimates and the mean state holding times from the Viterbi algorithm (in preference to using the diagonal generator elements), as described in section~\ref{constructionofHMM}.  The volumes of arriving fluid of each type in each bin are proportional to the numbers of reads and writes in these bins, the constant of proportionality being chosen as described in~\cite{peva10}.  Finally, the fluid arrival rate in each hidden state can then be obtained from the state's observation probabilities (in the $G$ matrix) or by inspection of the bins associated with that state in the Viterbi trace; we chose the latter option.

The outputs of the performance model include the probability densities of the fluid level at equilibrium and the moments of the queueing time for each type of IO operation.  A bin size of five milliseconds was again found to be suitable: too small a time
interval created too many empty intervals, whereas too large an interval missed crucial mode transitions.  

\subsection{Fluid queue specification}
Fluid models can be tailored to represent the behaviour of Flash memory executing different access
 operations, namely erase, write and read, because they can take into account correlation
 between access streams as well as  their relative priorities.  The modulating Markov
 chain is used to describe the evolution of the access mode, including any transformations induced by a file management system such as WAFL.
Specifically, we use a fluid queue with input defined by a four-phase continuous time semi-Markov chain (CTSMC); three phases represent different mixes of access modes and the fourth corresponds to erase operations.  Such a queue is particularly  appropriate (but not exclusively so) at moderate to high utilisation.  In our model, all phase holding times are assumed to be negative exponential random variables so that the modulating process is a CTMC.

The four phases identified by the HMM correspond to \emph{primarily read requests}, \emph{large write requests} (combined with some reads), \emph{small write requests} (also combined with some reads) and \emph{erases}.  Fluid ``particles'' arriving in read-mode have priority over the other modes to account for the priority actually given to reads, which are usually more critical in allowing a process to continue.  Fluid is output by the server at a constant rate when the buffer is not empty, and at zero rate when it is.  The different sizes of read and write requests are accounted for by assigning different volumes of fluid to each type of IO operation.  This model showed promise in its predicted queueing time probability densities when its parameters were estimated according to ``educated guesses'' in~\cite{peva10}.  However, it is important to choose its parameters in a quantitatively reliable way, and this is what clustering of the input data into a finite number of meaningful observation classes, followed by application of the HMM, has successfully provided. 
It is also important to map the discrete spatial behaviour of the storage system modelled onto the continuous state space (fluid) model.  This is done by defining the notion of \emph{vbytes} (virtual bytes), a real number that represents the completion time of one particular IO operation (here, a page-read); see \cite{peva10}. 

The fluid model determines the Laplace transform of the probability
density of the queueing time for arrivals in each phase at the chip.
For the preemptive read, high priority mode:
\begin{equation} \label{q1}
Q_H^*(\theta) = L_H^*(\theta/\mu)
\end{equation}
where $L_H^*(\theta)$ is the Laplace transform of the equilibrium read-fluid level density function at the arrival instant of a high priority fluid particle (this is a standard result, e.g.~\cite{mitra88}), and $\mu$ 
is the rate of the server.  The equation is more complex for the low priority classes but this is not the focus of the present paper;  it may be found in~\cite{peva10}, wherein typical graphs for the corresponding densities are shown.

\subsection{Flash performance: queueing times}
Considering the $G$ matrix and the clusters' centroids, we see that, to
parameterize the CTMC controlling the fluid input:
\begin{itemize}
\item Hidden state 1 is mostly issuing reads, so can be
  considered the ``READ'' state;
\item Hidden state 2 issues writes of all sizes at relatively high intensity, along with some reads: it is taken to be the ``LARGE WRITE'' state;
\item Hidden state 3 issues a mixture of writes and 
  reads, and is considered the ``WRITE-READ'' state;
\item Hidden state 4 is constructed so as to be the ``ERASE'' state, by making transition probabilities to it 1/64 of the existing transition probabilities to the LARGE WRITE state (and renormalising), and all transitions from it going to the LARGE WRITE state with probability 1.
\end{itemize}
Of course, the parameterisation of state 4 is somewhat arbitrary here; essentially, erases were added to match the amount of page writes, consistent with the scheduling of the FTL and higher level file system such as WAFL.  This matching was used similarly to estimate the mean holding time in the ERASE state, also allowing for the larger erase blocks.  In a real application, of course, more explicit 
account would be taken of these higher levels.
The CTMC thus constructed was used to modulate the fluid-input rate to the Flash model.  As observed in the IO traces, the write blocks arrive in an on-off manner, in practice because write traffic is de-staged lazily by the file system. The other hidden states appear to correspond to higher levels of write activity, combined with a largely consistent read pattern.  

\begin{figure}[!h]
\setlength{\epsfxsize}{0.7 \hsize}
\centerline{\epsfbox{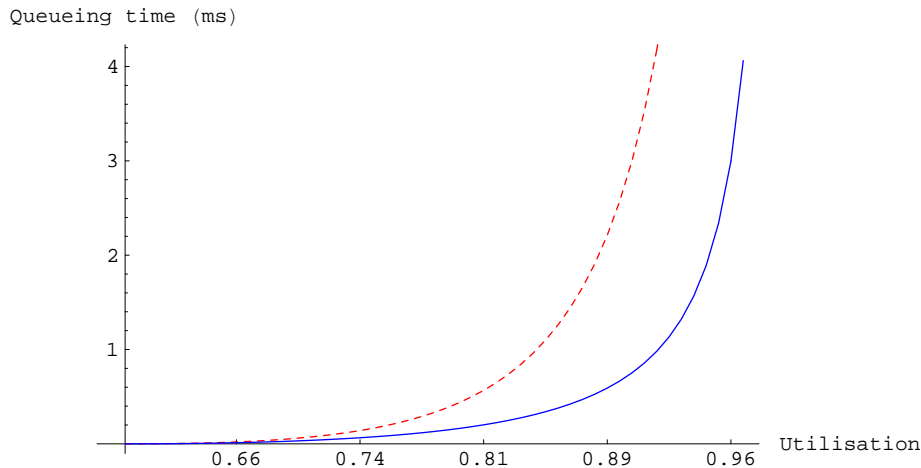}}
\caption{\label{qt} Mean queueing time for the aggregate low priority class in the \emph{oltp-mix} workload.  The dashed (red) curve corresponds to the preemptive priority scheme for reads and the solid (blue) curve corresponds to no priority.}
\end{figure}

For the first, write-dominated, \emph{update-mix}  workload considered, the read input rate is less than the service rate.  Hence, the preemptive, high priority reads never have to queue; so the level of high priority fluid must always be zero. 
We therefore focus on the low priority class representing writes and erases, and investigate the penalty these access modes suffer due to the read traffic.    Note that with the \emph{oltp-mix} workload, where reads dominate and increase the fluid level on their own (read-fluid input rate is greater than the fluid output rate), this penalty is much greater than for the \emph{update-mix}.
Given the model parameterisation corresponding to the workload trace analysed in section~\ref{tests}, and using a fluid service rate of $157.5$ (see~\cite{peva10} for justification of this and other rates), Figure~\ref{qt} shows the penalty on the low priority classes for the \emph{oltp-mix} workload.  It compares the respective mean queueing times for the low priority aggregate class (mixture of erases, large writes and read-writes) in the no-priority and preemptive read priority scenarios, as the utilisation of the chip increases.


For the \emph{update-mix} workload, the graphs are less dramatic with only a very small penalty, for the reason explained above.  In fact the relative maximum penalty -- at 98\% utilisation -- is just 6\%.

\begin{figure}[!h]
\setlength{\epsfxsize}{0.7 \hsize}
\centerline{\epsfbox{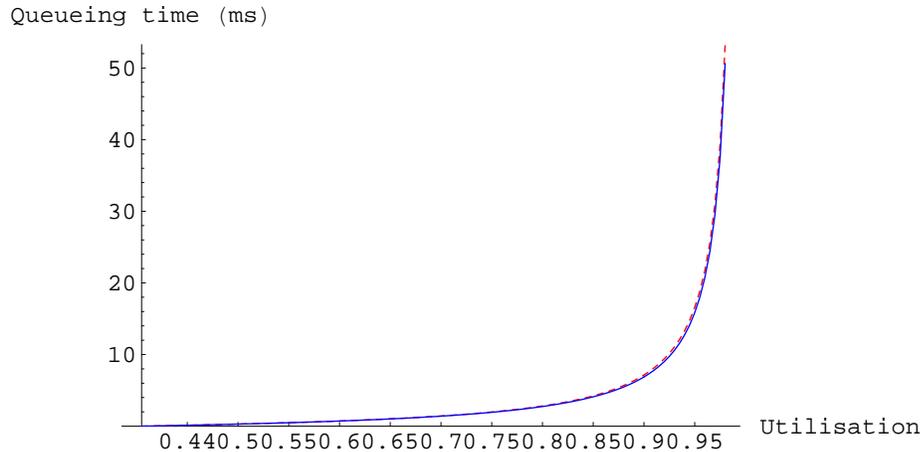}}
\caption{\label{qtupd} Mean queueing time for the aggregate low priority class in the \emph{update-mix} workload.  The dashed (red) curve corresponds to the preemptive priority scheme for reads and the solid (blue) curve corresponds to no priority.}
\end{figure}

\section{Conclusion and future work}
\label{concl}

We have described a HMM-based methodology for deriving a concise, parsimonious and portable synthetic workload benchmark from SPC-1 IO traces.   In particular, we have shown that the HMM trace metrics validate very well with the trace that is given as input. From the validated HMM model, we can create a MAP workload model that can be used as input to other performance models. The compact representation of the MAP (transition probability matrix, mean state holding times, and operation mix generation rates) captures important quantitative characteristics of the arrival streams, not possible with simpler abstractions like Poisson processes.
This type of workload characterization is important for Flash models
because correlation within and between the read and write streams is significant, perhaps
more so than the randomness of the accessed locations, which is a
better known distinguishing feature.

Mathematical descriptions of workload of the kind we have obtained must ultimately be assessed quantitatively against independent traces (i.e. not used in model construction) that they purport to represent.  The suit of statistical tests for goodness of fit used in section~\ref{tests} will be applied to such new data, characterised solely by its SPC-1 (or other benchmark) parameters in the case of another synthetic workload or by the arrival rates and block size distributions in the case of a production workload.  Further tests could also be applied, such as the partial ACF (PACF) and power spectrum.

There are many directions in which this work can be further extended.  Currently we are considering other levels in the storage stack and  the possibility of handling infrequent, higher intensity, additional loads.  Ideally, we would like to use our methodology for on-line characterization of workloads.  This could be achieved through  an \emph{incremental} HMM, i.e. one that has its parameters updated on-the-fly as more real time workload data becomes available.  It is not easy to devise such a model, but without the incremental feature, continually generating new HMMs at run-time would require a separate, dedicated processor -- maybe a possibility these days.

Finally, we also intend to improve
the fluid model by taking into account  both the non-preemptive
execution mode, which will significantly affect the performance of the
system in terms of response time,  and  non-Poisson arrival
streams.

\bibliographystyle{plain}
\bibliography{FlashHMM_peva}

\appendix 
\section{The EM algorithm for HMM}\label{EMHMM}
Let $(C_t, S_t)$ be a HMM with
$$P(C_t=c_t, S_t=s_t \mid C_{t-1}= c_{t-1},  S_{t-1}=s_{t-1})$$
$$P(C_t=c_t, S_t=s_t \mid C_{t-1}=q_{c_{t-1}c_t}(\theta) g_{c_ts_t}(\theta)$$
where $\theta$ is a vector of unknown parameters. We assume that the initial distribution $\nu$ is either known or fully determined by the parameter $\theta$. We assume that the observations $s_0, \ldots, s_n$ are available. The complete data likelihood is
$$J_{\nu,n} (s_0^n, c_0^n, \theta)=\nu_{c_0}(\theta) g_{c_0s_0}(\theta) \prod_{i=1}^n q_{c_{i-1}c_i}(\theta) g_{c_is_i}(\theta)$$
and so, 
$$\log J_{\nu,n} (s_0^n, c_0^n) = \log \nu_{c_0}( \theta) + \sum_{i=0}^{n-1} \log q_{c_{i}c_{i+1}}(\theta) + \sum_{i=0}^{n} \log g_{c_is_i}(\theta)$$
and the intermediate quantity of EM is
\begin{eqnarray}
\cal Q(\theta; \theta') &=& E_{\theta'} [\log J_{\nu,n} (s_0^n, C_0^n) \mid S_0^n=s_0^n] \nonumber \\ 
&=&\sum_{c_0} \log \nu_{c_0}(\theta) \phi_{0 \mid
  n}(c_0;\theta')
\nonumber \\ 
&& + \sum_{i=0}^{n-1} \sum_{c_i, c_{i+1}} \log
q_{c_ic_{i+1}}(\theta) \phi_{i : i+1 \mid n}(c_i, c_{i+1}; \theta')
\nonumber \\
&& +\sum_{i=0}^{n} \sum_{c_i} \log g_{c_is_i}(\theta) \phi_{i \mid n}(c_i, \theta') \label{calQ}
\end{eqnarray}





\begin{proposition} \label{emparameters}
The parameters that maximise $Q(\theta; \theta')$~are
$$ \hat q_{jk}= \frac{\sum_{i=0}^{n-1}\phi_{i : i+1 \mid n}(j, k~; \theta')} {\sum_{i=0}^{n-1} \phi_{i \mid n}(j~; \theta')}$$
$$\hat g_{js}= \frac{ \sum_{i=0}^{n} \delta_{s_i,s}  \phi_{i \mid n}(j; \theta')}{\sum_{i=0}^{n}  \phi_{i \mid n}(j; \theta')}$$
$$\hat \nu_j = \frac{  \phi_{0 \mid n}(j; \theta')} {\sum_{l}   \phi_{0 \mid n}(l; \theta')}$$
\end{proposition}
\pf
First, consider the second term in equation \ref{calQ}, say:
\begin{eqnarray*}
F_2=\sum_{i=0}^{n-1} \sum_{j, k} \log q_{jk}(\theta) \phi_{i : i+1 \mid n}(j, k; \theta')
\end{eqnarray*}
Adding the Lagrange multiplier $\mu$, using the constraint that
$\sum_{l} q_{jl} =1$, and setting the derivative to zero we have\\
$\frac{\partial}{\partial q_{jk}}[F_2+\mu \sum_{l}q_{j,k}]=\sum_{i=0}^{n-1}\frac{1}{q_{jk}}\phi_{i:i+1\mid n}(j, k;\theta')+\mu=0$\\\\
So\\
$$\hat q_{jk} = -\frac{1}{\mu} \sum_{i=0}^{n-1} \phi_{i : i+1 \mid
  n}(j, k; \theta')$$
And by the constraint\\
$$1 = -\frac{1}{\mu} \sum_{l} \sum_{i=0}^{n-1}  \phi_{i : i+1 \mid
  n}(j, l; \theta') $$
So\\
$$\mu = - \sum_{i=0}^{n-1} \phi_{i \mid n}(j; \theta') $$
Therefore\\
$$\hat q_{jk} = \frac{\sum_{i=0}^{n-1}\phi_{i : i+1 \mid n}(j, k~;
  \theta')} {\sum_{i=0}^{n-1} \phi_{i \mid n}(j~; \theta')}$$

\medskip \noindent
Next, consider the third term in equation \ref{calQ}:
\begin{eqnarray*}
F_3=\sum_{i=0}^{n} \sum_{j} \log g_{js_i}(\theta) \phi_{i \mid n}(j, \theta') 
  \end{eqnarray*}
 Adding the Lagrange multiplier $\mu$, using the constraint that $\sum_{s} g_{ls}=1$ for all $l \in {\cal S}$, and setting the derivative to zero we have
 \begin{eqnarray*}
 \frac{\partial}{\partial g_{js}} [ F_3+\mu \sum_{s} g_{l,s} ]&=&\sum_{i=0}^{n} \frac{\delta_{s_i,s}}{g_{js}}  \phi_{i \mid n}(j, \theta') + \mu =0\\
 \mbox{So~~~~~~~~~~~~~~~~~~~~~~~~~~~~~} \hat g_{js}&=& -\frac{1}{\mu} \sum_{i=0}^{n} \delta_{s_i,s}  \phi_{i \mid n}(j, \theta')\\
  \mbox{ And by the constraint~~~~~~} 1 &=& -\frac{1}{\mu} \sum_{s}   \sum_{i=0}^{n}  \delta_{s_i,s}  \phi_{i \mid n}(j, \theta')\\
   \mbox{So~~~~~~~~~~~~~~~~~~~~~~~~~~~~~~~} \mu &=& - \sum_{i=0}^{n} \phi_{i \mid n}(j, \theta')\\
   \mbox{Therefore~~~~~~~~~~~~~~~~~~~~~~~~~~~}  \hat g_{js}&=& \frac{ \sum_{i=0}^{n} \delta_{s_i,s}  \phi_{i \mid n}(j, \theta')}{\sum_{i=0}^{n}  \phi_{i \mid n}(j, \theta')}
   \end{eqnarray*}

\medskip \noindent
Finally, consider the first term in equation \ref{calQ}:
\begin{eqnarray*}
F_1=\sum_{c_0} \log \nu_{c_0}(\theta) \phi_{0 \mid n}(c_0;\theta')
\end{eqnarray*}
Adding the Lagrange multiplier $\mu$, using the constraint that $\sum_{l} \nu_{l} =1$, and setting the derivative to zero we have
\begin{eqnarray*}
 \frac{\partial}{\partial \nu_{j}} [ F_1+\mu \sum_{l} \nu_{l} ]&=& \frac{1}{\nu_j} \phi_{0 \mid n}(j; \theta') + \mu =0\\
 \mbox{So~~~~~~~~~~~~~~~~~~~~~~~~~~~~~} \hat \nu_j &=& -\frac{1}{\mu} \phi_{0 \mid n}(j; \theta')\\
 \mbox{ And by the constraint~~~~~~} 1 &=& -\frac{1}{\mu} \sum_{l}  \phi_{0 \mid n}(l; \theta')\\
 \mbox{So~~~~~~~~~~~~~~~~~~~~~~~~~~~~~~} \mu &=& -  \sum_{l}   \phi_{0 \mid n}(l; \theta')\\
 \mbox{Therefore~~~~~~~~~~~~~~~~~~~~~~~~~~~}  \hat \nu_j &=& \frac{  \phi_{0 \mid n}(j; \theta')} {\sum_{l}   \phi_{0 \mid n}(l; \theta')} \qquad \diamondsuit
 \end{eqnarray*}

\section{HMM for CIFS data showing cyclic behaviour} \label{B}
A 12-state HMM was obtained for the CIFS observation trace used in section~\ref{cifs100ms} in order to check whether any HMM could describe the periodic behaviour observed in the raw data.  This only required the number of hidden states being set to 12 in the Baum-Welch algorithm, leading to an increase in computation time by a factor of approximately $(12/4)^2 = 9$, according to the result in section~\ref{constructionofHMM}.  Actually, the factor was nearer to 5 due to a decrease in the number of iterations required to reach the desired precision.   We obtained the following parameter estimates. 

For the one-step transition probability matrix of the HMM, 

\medskip
{\footnotesize $\hat Q =$}
{\tiny
$$\hspace{-2cm} \left(
\begin{array}{cccccccccccc}
 0 & 0 & 0 & 0 & 0 & 0 & 0 & 0 & 1. & 0 & 0 & 0 \\
 0 & 0 & 0 & 0 & 0 & 0 & 0 & 0 & 0 & 0 & 0 & 1. \\
 1.47\; 10^{-5} & 4.25\; 10^{-8} & 9.91\; 10^{-1} & 2.8\; 10^{-4} & 2.72\; 10^{-8} & 0 & 2.52\; 10^{-5} & 7.94\; 10^{-3} & 0 &
   1.95\; 10^{-6} & 8.34\; 10^{-4} & 5.99\; 10^{-8} \\
 0 & 0 & 7.07\; 10^{-1} & 3.88\; 10^{-3} & 2.89\; 10^{-1} & 0 & 0 & 5.23\; 10^{-8} & 0 & 5.48\; 10^{-6} & 8.87\; 10^{-9} & 0 \\
 0 & 0 & 4.73\; 10^{-2} & 7.1\; 10^{-7} & 0 & 9.53\; 10^{-1} & 0 & 0 & 0 & 0 & 0 & 0 \\
 0 & 1. & 5.26\; 10^{-6} & 5.74\; 10^{-10} & 0 & 0 & 0 & 0 & 4.33\; 10^{-7} & 0 & 0 & 0 \\
 0 & 0 & 8.05\; 10^{-5} & 3.52\; 10^{-3} & 0 & 0 & 0 & 0 & 0 & 9.96\; 10^{-1} & 0 & 0 \\
 0 & 0 & 2.82\; 10^{-10} & 0 & 0 & 0 & 0 & 0 & 0 & 0 & 1. & 0 \\
 0 & 0 & 0 & 0 & 0 & 0 & 0 & 1. & 0 & 0 & 0 & 0 \\
 0 & 0 & 4.87\; 10^{-3} & 2.82\; 10^{-8} & 9.95\; 10^{-1} & 0 & 0 & 0 & 0 & 0 & 0 & 0 \\
 0 & 0 & 6.92\; 10^{-8} & 2.04\; 10^{-5} & 0 & 0 & 1. & 0 & 0 & 0 & 0 & 0 \\
 9.99\; 10^{-1} & 9.41\; 10^{-4} & 0 & 0 & 0 & 0 & 0 & 0 & 0 & 0 & 0 & 0
\end{array}
\right)$$
}
The observation values in each state had probabilities estimated by the matrix $\hat G$, written as its transpose for the sake of readability on the printed page,
{\footnotesize 
$$\hat G^T = \left(
\begin{array}{cccccccccccc}
 1. & 4.38\; 10^{-1} & 9.87\; 10^{-1} & 1. & 1. & 4.44\; 10^{-1} & 1. & 1. & 6.51\; 10^{-2} & 1. & 1. & 5.06\; 10^{-1} \\
 0 & 2.83\; 10^{-1} & 1.03\; 10^{-2} & 0 & 0 & 2.98\; 10^{-1} & 0 & 0 & 7.55\; 10^{-2} & 0 & 0 & 3.16\; 10^{-1} \\
 0 & 3.72\; 10^{-3} & 0 & 0 & 0 & 5.58\; 10^{-3} & 0 & 0 & 2.79\; 10^{-3} & 0 & 0 & 1.86\; 10^{-3} \\
 0 & 1.39\; 10^{-2} & 1.51\; 10^{-4} & 0 & 0 & 1.67\; 10^{-2} & 0 & 0 & 4.83\; 10^{-2} & 0 & 0 & 1.3\; 10^{-2} \\
 0 & 0 & 2.88\; 10^{-4} & 0 & 0 & 9.3\; 10^{-4} & 0 & 0 & 0 & 0 & 0 & 0 \\
 0 & 2.49\; 10^{-1} & 1.82\; 10^{-3} & 0 & 0 & 2.25\; 10^{-1} & 0 & 0 & 1.01\; 10^{-1} & 0 & 0 & 1.59\; 10^{-1} \\
 0 & 4.25\; 10^{-3} & 2.06\; 10^{-4} & 0 & 0 & 3.72\; 10^{-3} & 0 & 0 & 2.79\; 10^{-3} & 0 & 0 & 9.28\; 10^{-4} \\
 0 & 9.29\; 10^{-4} & 0 & 0 & 0 & 2.79\; 10^{-3} & 0 & 0 & 1.02\; 10^{-2} & 0 & 0 & 0 \\
 0 & 0 & 0 & 0 & 0 & 0 & 0 & 0 & 1.86\; 10^{-3} & 0 & 0 & 0 \\
 0 & 6.25\; 10^{-3} & 0 & 0 & 0 & 1.86\; 10^{-3} & 0 & 0 & 3.52\; 10^{-1} & 0 & 0 & 2.11\; 10^{-3} \\
 0 & 0 & 1.37\; 10^{-4} & 0 & 0 & 9.3\; 10^{-4} & 0 & 0 & 1.21\; 10^{-2} & 0 & 0 & 9.29\; 10^{-4} \\
 0 & 0 & 0 & 0 & 0 & 0 & 0 & 0 & 4.09\; 10^{-2} & 0 & 0 & 0 \\
 0 & 9.3\; 10^{-4} & 0 & 0 & 0 & 0 & 0 & 0 & 2.53\; 10^{-1} & 0 & 0 & 0 \\
 0 & 0 & 0 & 0 & 0 & 0 & 0 & 0 & 8.36\; 10^{-3} & 0 & 0 & 0 \\
 0 & 0 & 0 & 0 & 0 & 0 & 0 & 0 & 2.6\; 10^{-2} & 0 & 0 & 0
\end{array}
\right)$$
}
Finally, the initial probability estimates were
{\footnotesize 
$$\hat \nu = \left(
\begin{array}{cccccccccccc}
 0 & 0 & 0 & 0 & 0 & 0 & 0 & 0 & 0 & 0 & 0 & 1.
\end{array}
\right)$$
}
Using the same procedure as in section~\ref{cifs100ms}, we obtained the following ACF for read operations from the HMM-generated traces, plotted for lags up to 91.  

\begin{figure}[!h]
\centerline{
\setlength{\epsfxsize}{0.7 \hsize}
\epsfbox{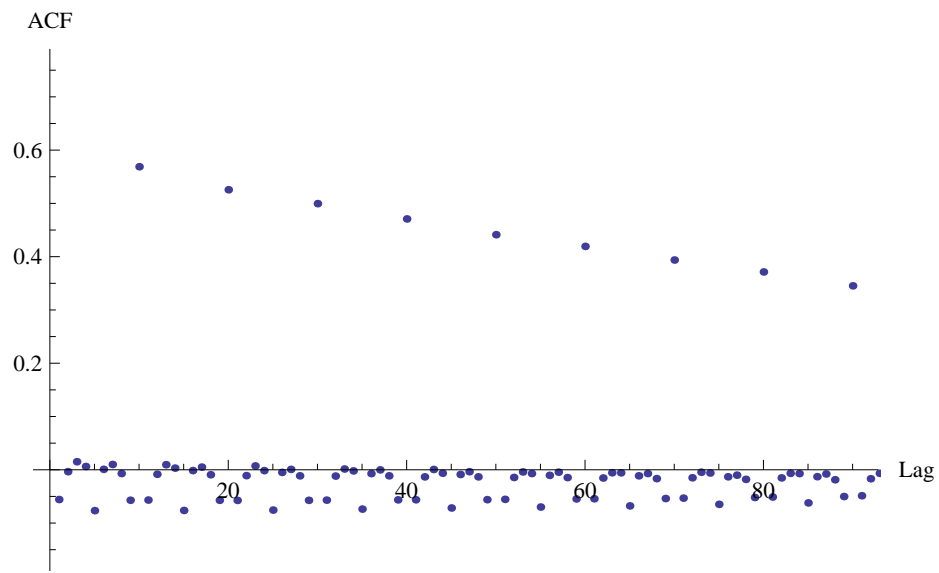}}
\caption{\label{CIFSHMM100ms_0-91}  Exploded view of the ACF for reads in the 12-state HMM-generated100ms binned traces. }
\end{figure}

This graph is in excellent agreement with that of Figure~\ref{CIFS100ms_0-91}.  The corresponding ACF for write operations also shows the correct periodicity, 10, albeit with a smaller amplitude than seen in Figure~\ref{CIFS100ms_0-91}.  We believe this to be because the writes are dominated by the reads in the CIFS time series -- there are 93\% reads -- so that a longer series would be necessary to estimate the ACF, in both cases of the HMM and raw traces.

\end{document}